\documentclass[12pt]{article}
\pdfoutput=1

\usepackage{tikz} 

\usepackage[utf8]{inputenc}
\usepackage[left=2.55cm, right=2.55cm, top=2.55cm, bottom=2.55cm]{geometry}
\usepackage{amsmath,amssymb,amsbsy}
\usepackage{slashed}
\usepackage{xcolor}
\usepackage{graphicx}
\usepackage{url}
\usepackage{cancel}
\usepackage{cite}
\usepackage[colorlinks=true,allcolors=blue,pdfborder={0 0 0},linktocpage=false,pdfencoding=auto]{hyperref}
\usepackage{tabularx,booktabs}
\usepackage{multicol}
\usepackage{feynmp}
\usepackage{units}
\usepackage{xspace}
\usepackage[labelfont=bf]{caption}
\usepackage[section]{placeins}
\usepackage{subcaption}
\usepackage{soul}
\usepackage{bbold}
\usepackage{parskip}
\usepackage{float}
\usepackage{tabulary}
\usepackage{bm}

\definecolor{darkred}{rgb}{0.6,0,0}
\definecolor{darkpurple}{rgb}{0.5,0,0.5}

\newcommand{\code}[1]{\texttt{#1}}
\newcommand{\beqn}{\begin{eqnarray}}
\newcommand{\eeqn}{\end{eqnarray}}
\def\non{\nonumber\\}

\newcommand{\cl}{{\cal L}}
\def\stu{Stueckelberg~}

\newcommand{\swatch}[1]{\tikz[baseline=-0.6ex]\node[fill=#1,shape=rectangle,draw=black,thick,minimum width=5mm,rounded corners=0.5pt](){};}
\newcommand{\met}{\ensuremath{E_T^{\rm miss}}}

\definecolor{powderblue}{HTML}{B0E0E6}
\definecolor{green}{HTML}{008000}
\definecolor{seagreen}{HTML}{2E8B57}
\definecolor{orangered}{HTML}{FF4500}
\definecolor{orange}{HTML}{FFA500}
\definecolor{darkolivegreen}{HTML}{556B2F}
\definecolor{mediumseagreen}{HTML}{3CB371}
\definecolor{darksalmon}{HTML}{E9967A}
\definecolor{salmon}{HTML}{FA8072}
\definecolor{darkorange}{HTML}{FF8C00}
\definecolor{silver}{HTML}{C0C0C0}
\definecolor{dimgrey}{HTML}{696969}
\definecolor{yellow}{HTML}{FFFF00}
\definecolor{tomato}{HTML}{FF6347}
\definecolor{cornflowerblue}{HTML}{6495ED}
\definecolor{cadetblue}{HTML}{5F9EA0}
\definecolor{turquoise}{HTML}{40E0D0}
\definecolor{darkorchid}{HTML}{9932CC}
\definecolor{darkgoldenrod}{HTML}{B8860B}
\definecolor{navy}{HTML}{000080}
\definecolor{magenta}{HTML}{FF00FF}
\definecolor{indigo}{HTML}{4B0082}

\begin{document}

\author{Amin Aboubrahim$^a$\footnote{\href{mailto:aabouibr@uni-muenster.de}{aabouibr@uni-muenster.de}}~, Mohammad Mahdi Altakach$^b$\footnote{\href{mailto:mohammadmahdi.takach@live.com}{altakach@lpsc.in2p3.fr}}~, Michael Klasen$^{a,c}$\footnote{\href{mailto:michael.klasen@uni-muenster.de}{michael.klasen@uni-muenster.de}}~, \\
Pran Nath$^d$\footnote{\href{mailto:p.nath@northeastern.edu}{p.nath@northeastern.edu}}~ and Zhu-Yao Wang$^d$\footnote{\href{mailto:wang.zhu@northeastern.edu}{wang.zhu@northeastern.edu}} \\~\\
$^{a}$\textit{\normalsize Institut f\"ur Theoretische Physik, Westf\"alische Wilhelms-Universit\"at M\"unster,} \\
\textit{\normalsize Wilhelm-Klemm-Stra{\ss}e 9, 48149 M\"unster, Germany} \\
$^{b}$\textit{\normalsize Laboratoire de Physique Subatomique et de Cosmologie, CNRS/IN2P3,}\\
\textit{\normalsize 53 Avenue des Martyrs, 38026 Grenoble, France} \\
$^{c}$\textit{\normalsize School of Physics, The University of New South Wales, Sydney NSW 2052, Australia}\\
$^{d}$\textit{\normalsize Department of Physics, Northeastern University,} \\
\textit{\normalsize 111 Forsyth Street, Boston, MA 02115-5000, USA}
}

\title{\vspace{-3cm}\begin{flushright}
{\small MS-TP-22-50}
\end{flushright}
\vspace{1cm}
\Large \bf
Combined constraints on dark photons and discovery prospects at the LHC and the Forward Physics Facility
\vspace{0.5cm}}

\date{}
\maketitle

\vspace{1cm}

\begin{abstract}

Hidden sectors are ubiquitous in supergravity theories, in strings and in branes. Well motivated models such as the Stueckelberg hidden sector model could provide a  candidate for dark matter. In such  models, the  hidden sector communicates with the visible sector via the exchange of a dark photon (dark $Z'$) while dark matter
is constituted of Dirac fermions in the hidden sector.
Using data from collider searches and precision measurements of SM processes as well as the most recent limits from dark matter direct and indirect detection experiments, we perform a comprehensive scan over a wide range of the $Z'$ mass and set exclusion bounds on the parameter space from sub-GeV to several TeV. We then discuss the discovery potential of an $\mathcal{O}$(TeV) scale $Z'$ at HL-LHC and the ability of future forward detectors to probe very weakly interacting sub-GeV $Z'$ bosons. Our analysis shows that the parameter space in which a $Z'$ can decay to hidden sector dark matter is severely constrained whereas limits become much weaker for a $Z'$ with no dark decays. The analysis also favors a self-thermalized dark sector which is necessary to satisfy the dark matter relic density.     

\end{abstract}

\numberwithin{equation}{section}
\newpage

{  \hrule height 0.4mm \hypersetup{colorlinks=black,linktocpage=true} \tableofcontents
\vspace{0.5cm}
 \hrule height 0.4mm}

\section{Introduction}

The identification of dark matter (DM) is one of the most important current problems in astroparticle physics, given the overwhelming observational evidence for it on many different length scales. Aside from visible sector candidates for DM such as the  weakly interacting massive particle (WIMP),  one may have dark matter arising from the hidden sector. Here we discuss experimental limits on a model containing such a candidate that arises from the extension of the electroweak sector of the Standard Model (SM) along with an additional $U(1)_X$ gauge group of the hidden sector which kinetically mixes with the gauge field of the hypercharge  $U(1)_Y$~\cite{Holdom:1985ag,Feldman:2007wj}. For this, we consider a well motivated  extension with the Stueckelberg mechanism as the source of mass generation for the extra gauge boson. As noted, the communication between the hidden and the visible sectors arises due to kinetic mixing of $U(1)_X$ and $U(1)_Y$. In this case, transition to the canonical basis requires diagonalization of a $3\times 3$ mass matrix for the gauge fields involving $U(1)_X, U(1)_Y$ and the 
gauge field for the neutral component of the group $SU(2)_L$. In the canonical basis one finds one massless field which is the photon and two massive fields which can be identified as the $Z$ boson
and a massive dark field. This massive dark field can be viewed as a dark photon, $Z'$, since it is associated
with a $U(1)_X$ factor and mixes with the SM photon. In this work, $Z'$ will take on a wide range of masses, from sub-GeV to multi-TeV values.
In the canonical basis, $Z'$ will have interactions with the visible sector quarks and leptons
while the $Z$ boson will also have interactions with the dark fermions which carry $U(1)_X$ quantum 
numbers. Thus communication exists between the hidden and the visible sectors due to the exchange
of $Z'$ and the exchange of $Z$. Several recent works in this framework can be found in~\cite{Aboubrahim:2020lnr,Aboubrahim:2020afx,Aboubrahim:2021ycj,Aboubrahim:2021ohe}. For an overview of heavy $Z'$ models see ref.~\cite{Langacker:2008yv} and of dark photons see ref.~\cite{Fabbrichesi:2020wbt}.

We will discuss the \stu model in more detail in section~\ref{sec:model}. However, here we give a brief review of the mechanism for easy reference. Thus, let us consider the Lagrangian with an abelian vector
boson $C_\mu$  coupled to a pseudo-scalar $\sigma$ so that
\beqn
{\cal L} = -\frac14 { C}_{\mu\nu}{ C}^{\mu\nu} - \frac12
(m C_\mu + \partial_\mu \sigma)(m C^\mu + \partial^\mu \sigma)\ .
\eeqn
This Lagrangian is  gauge invariant under the transformations
$\delta C_\mu = \partial_\mu \epsilon$ and $\delta \sigma = - m \epsilon.$
Using the gauge fixing term 
$\cl_{\rm gf} = - \frac{1}{2\xi} \left( \partial_\mu A^\mu + \xi m \sigma \right)^2$,
the resulting Lagrangian has the form
\beqn
\cl_{\rm tot}=
-\frac14 {C}_{\mu\nu}{C}^{\mu\nu} - \frac{m^2}{2} C_\mu C^\mu - \frac{1}{2\xi} (\partial_\mu C^\mu)^2
- \frac12 \partial_\mu \sigma \partial^\mu \sigma - \xi \frac{m^2}{2} \sigma^2.
\eeqn
Here one finds that $C_\mu$ is massive and decoupled from $\sigma$.
We can include couplings with matter in the usual way using the
interaction term
${\cal L}_{\rm int} = g A_\mu J^\mu$ along with the constraint $\partial_\mu J^\mu = 0$.

As noted earlier, in this work we will consider a \stu extension of the SM.
In the analysis, we use the coupling that $Z'$ has with the SM fermions to derive constraints on the model parameters based on searches at the LHC as well as precision measurements of SM processes using \code{Contur}, a new tool to set exclusion limits. Further, DM searches from direct and indirect detection experiments add more constraints which also depend on the coupling between $Z'$ and the dark fermions. The analysis covers a wide range of the $Z'$ mass, from the sub-GeV 
to multi-TeV mass range, and we discuss the parameter space remaining after the multitude of constraints from a variety of experiments have been imposed. We also perform a detailed analysis for a potential discovery of $Z'$ at the high luminosity LHC (HL-LHC) and discuss the mass reach of future forward detectors in the sub-GeV mass region of the $Z'$ boson. Several precision calculations have been performed for LHC production of a heavy $Z'$~\cite{Fuks:2007gk,Klasen:2016qux} and for $t\bar{t}$ production with $Z'$ and $W'$~\cite{Bonciani:2015hgv,Altakach:2020ugg,Jezo:2014wra}. In this work, our calculation is done at the NLO level without including the resummation effects.   

The outline of the remainder of the paper is as follows: In section~\ref{sec:model} we define the hidden sector model and its communication with the visible sector and specifically of the $Z'$ with the Standard Model particles and with the dark fermion in the hidden sector. In section~\ref{sec:experiment} we list the different collider and dark matter direct and indirect detection experiments whose limits are recasted and used to constrain our model which we show in section~\ref{sec:limits}.  In section~\ref{sec:lhc} we give a detailed LHC analysis for a potential discovery of a TeV mass $Z'$. 
The sensitivity reach at forward detectors for a sub-GeV dark photon is discussed in section~\ref{sec:forward}.
Conclusions are given in section~\ref{sec:conclusion}. Further details related to the model are given in Appendix~\ref{app:A} while exclusion limits from \code{Contur} are shown in Appendix~\ref{app:C}. 

\section{Dark photons in the Stueckelberg extension of the Standard Model}\label{sec:model}

We give now a brief account of the extension of the electroweak sector of the Standard Model with an extra $U(1)_X$. The gauge 
content and matter content of this sector consist of the gauge field $C_\mu$ and the dark Dirac fermion $D$. We will assume a kinetic mixing of the $U(1)_X$ gauge field with the hypercharge gauge field of the SM as well as a \stu mass growth of the $U(1)_X$ gauge field and for generality we allow a mass mixing of the $U(1)_X$ gauge field with the $U(1)_Y$ field. Thus, the extended electroweak sector has the following Lagrangian~\cite{Kors:2004dx}
\begin{align}
\mathcal{L}=\mathcal{L}_{\rm SM}+\Delta \mathcal{L},
\label{totL}
\end{align}  
where $\Delta \mathcal{L}$ is the extended part of the Lagrangian given by 
\begin{align}
\Delta \mathcal{L} =&-\frac{1}{4}C_{\mu\nu}C^{\mu\nu}+i\bar D \gamma^\mu \partial_\mu D -m_D \bar D D \non
&-\frac{\delta}{2}C_{\mu\nu}B^{\mu\nu}-\frac{1}{2}(\partial_{\mu}\sigma+M_1 C_{\mu}+M_2 B_{\mu})^2
\non
 &+ g_X Q_X \bar D \gamma^\mu D C_\mu.  
\end{align}
Here the first line gives the free part of the Lagrangian for the $U(1)_X$ gauge boson $C^\mu$ and the hidden sector Dirac
fermion $D$, the second line gives the kinetic mixing and the 
\stu mass mixing for $C^\mu$ with the hypercharge gauge field
$B^\mu$, and the last line gives the interaction of $C^\mu$ 
with the Dirac fermion $D$.  To obtain the mass eigenstates, 
we diagonalize $\Delta \mathcal{L}$ along with the Standard
Model mass matrix for the gauge fields $B^\mu$ and $A^\mu_3$,
where the latter is the third component of the $SU(2)_L$ gauge field $A^\mu_a$ ($a=1,2,3$).  This requires diagonalizing a $3\times 3$ matrix involving the fields $C^\mu, B^\mu, A^\mu_3$ and in the eigen-frame where both the kinetic and the mass squared matrices are diagonalized, one has the gauge bosons $A^\mu_{\gamma}, Z^\mu, Z^{\prime\mu}$ corresponding to the particles $\gamma, Z, Z'$ where $\gamma$ is the photon, $Z$ is the $Z$-boson, and $Z'$ is the dark photon. In the canonically diagonalized frame the new interactions are given by   
\begin{align}
\label{D-darkphoton}
\Delta \mathcal{L}_{\rm int}&=\bar D\gamma^\mu(g_{Z'} Z^\prime_{\mu}+g_{Z} Z_{\mu}+g_{\gamma} A_{\mu}^{\gamma})D
   +  \frac{g_2}{2\cos\theta}\bar\psi_f\gamma^{\mu}(v'_f-\gamma_5 a'_f)Z'_{\mu}\psi_f.  
\end{align}
Here  $f$ stands for SM quarks and leptons. For the case when the kinetic mixing or \stu mass mixing is small, one has $g_{Z'}\simeq  g_X Q_X$, so $g_{Z'}$ is of normal size. It is given together with the couplings $g_{Z}$ and $g_{\gamma}$ in Appendix~\ref{app:A}. The dark photon can have vector and axial vector couplings with the fermions $f$ in the visible sector (quarks and leptons)
\begin{equation}
\begin{aligned}
v'_f&=-\cos\psi[(\tan\psi-s_\delta\sin\theta)T_{3f}-2\sin^2\theta(-s_{\delta} \csc\theta+\tan\psi)Q_f],\\
a'_f&=-\cos\psi(\tan\psi-s_{\delta} \sin\theta)T_{3f}.
\end{aligned}
\label{eqn:v-a1}
\end{equation}
Here $s_\delta= \sinh\delta$, $T_{3f}$ is the third component of isospin, $Q_f$ is the electric charge for the fermion $f$ and  $\theta$ and $\psi$ are angles defined in Appendix~\ref{app:A}.
Here we also note that the couplings of $Z_\mu$ and $A^\gamma_\mu$ in the canonically diagonalized basis  are also modified and are
given by~\cite{Feldman:2007wj}
\begin{equation}
\Delta \mathcal{L}'_{\rm SM}=\frac{g_2}{2\cos\theta}\bar\psi_f\gamma^{\mu}\Big[(v_f-\gamma_5 a_f)Z_{\mu}\Big]\psi_f+e\bar\psi_f\gamma^{\mu}Q_f A^\gamma_{\mu}\psi_f\,,
\label{SMLag}
\end{equation}
where the modifications appear in the definition of the vector coupling $v_f$ and the axial-vector coupling $a_f$ which are given by
\begin{equation}
\begin{aligned}
v_f&=\cos\psi[(1+ s_\delta \tan\psi\sin\theta)T_{3f}-2\sin^2\theta(1+ s_\delta \csc\theta\tan\psi)Q_f],\\
a_f&=\cos\psi(1 + s_\delta \tan\psi\sin\theta)T_{3f}. 
\end{aligned}
\label{eqn:v-a2}
\end{equation}  
Here Eqs.~(\ref{eqn:v-a1}) and~(\ref{eqn:v-a2}) are written with the assumption of zero mass mixing, i.e. $M_2=0$, which we take to be the case throughout this work.
We note in passing that in the literature the dark photon refers to a vector boson from an extra $U(1)$ gauge field that has kinetic mixing with a massless SM photon. In this case the dark photon does not mix with the $Z$ boson and so has no coupling to neutrinos. The \stu analysis is different in that here, the extra $U(1)$ gauge field mixes with the SM hypercharge $U(1)_Y$ gauge field and since $U(1)_Y$ gauge field has mixing with the gauge field for the neutral component of the $SU(2)_L$ triplet due to the SM Higgs mechanism one has in general a mass matrix which mixes the three neutral fields. The resulting massive $Z'$ has couplings to all SM fermions just as the $Z$ boson does. This is mentioned in order to draw a distinction between the two approaches. For related works and extensions see~\cite{Cheung:2007ut,Feldman:2006wd,Aboubrahim:2019kpb,Aboubrahim:2022bzk,Du:2022fqv}.

\section{Experimental constraints}\label{sec:experiment}

In this Section we list the different experiments and methods used in our analysis to constrain the parameter space of the Stueckelberg dark photon model. The experimental limits we use pertain to the $Z'$ as well as to the dark fermion DM which we turn into constraints on the model parameter space in the kinetic mixing-dark photon mass plane. The constraints are also sensitive to the $Z'$ decay channels, i.e., visible decays to SM fermions and invisible decays to the dark fermion which we will refer to as dark decays. 

The scan of the model parameter space and calculation of the relevant observables requires a tool chain which we describe next. The Stueckelberg extension of the SM with a dark sector is implemented in \code{SARAH}~\cite{Staub:2013tta,Staub:2015kfa} which produces the necessary \code{SPheno}~\cite{Porod:2003um,Porod:2011nf} model files for spectrum generation. \code{SARAH} also automatically writes the \code{CalcHep/CompHep}~\cite{Pukhov:2004ca,Boos:1994xb} files used by \code{micrOMEGAs}~\cite{Belanger:2014vza} and the \code{UFO}~\cite{Degrande:2011ua} files needed by \code{MadGraph5\_aMC@NLO}~\cite{Alwall:2014hca}. There are more necessary tools that have been used which we will refer to later in the paper.    

\subsection{Fiducial measurements of SM processes using \code{Contur}}

Hundreds of differential cross section measurements were performed during the first two runs of the LHC. These measurements, even though intended to test the SM of particle physics, can still be used to investigate physics beyond the SM (BSM) due to their model-independent nature. A software called 
\code{Contur} or “Constraints On New Theories Using Rivet”~\cite{Butterworth:2016sqg,Altakach:2021okx} scans the SM analyses implemented in the \code{Rivet} (Robust Independent Validation of Experiment and Theory) toolkit~\cite{Bierlich:2019rhm} to check whether a BSM signal is already excluded, and if so, at which significance. 
\code{Contur} performs a $\chi^2$ test statistic to evaluate the likelihood of the BSM model taking the experimental uncertainties into account. The considered hypotheses are SM-only and SM+BSM. The CL$_s$ technique~\cite{Junk:1999kv,Read:2002hq} is then used to derive the confidence level exclusion on the BSM theory at a given point of the parameter space. This procedure is then repeated for each set of parameter values resulting in a map of CL exclusions (more information about \code{Contur} and its statistical method used to perform the exclusion procedure can be found in~\cite{Buckley:2021neu}). Many studies were performed using the \code{Contur} toolkit to check different BSM scenarios~\cite{Butterworth:2022dkt,Altakach:2021lkq,Altakach:2022nuo,Butterworth:2020vnb} and it was shown that limits from precision measurements of SM processes can be more constraining than BSM searches at the LHC in some parts of the parameter space~\cite{Buckley:2020wzk}. In this work, we use \code{Contur} to set limits on the parameter space of our model. The signal cross section was calculated at leading order using \code{Herwig}~\cite{Bellm:2019zci} for a center of mass energy of $7$, $8$, and $13$ TeV. Details about the LHC analyses that contributed to the exclusions are given in Appendix~\ref{app:C}. We consider four cases in which the $Z'$ is either heavy or light, and whether it decays only to visible final states or to dark fermions as well. Furthermore, since $Z'$ mixes with $Z$, contributions to the $Z$ boson mass and width are expected, so additional LEP constraints are included~\cite{ALEPH:2005ab,Electroweak:2003ram,ParticleDataGroup:2018ovx}.

\subsection{LHC searches: dijet, dilepton and monojet limits}

We call $Z'$ heavy if its mass is greater than the $Z$ boson mass and light if it's smaller. 
Models with an extra neutral gauge boson $Z'$ are extensively tested and constrained by a myriad of collider searches. ATLAS and CMS experiments have searched for a heavy $Z'$ resonance and set stringent bounds on the ratio $m_{Z'}/g_X$, with $g_X$ being the gauge coupling for a particular $U(1)_X$ extension of the SM. Searches for light $Z'$ in the mass range of 1 to $\sim 80$ GeV have been carried out by CMS, LHCb and BaBar. Lighter masses have also been investigated at beam dump experiments. 

For heavy $Z'$, ATLAS and CMS collaborations have looked for an excess of events in the dijet invariant mass corresponding to the decay of a heavy vector resonance~\cite{CMS:2019gwf,ATLAS:2019fgd,ATLAS:2018qto,CDF:2008ieg,ATLAS:2018hbc,CMS:2019emo,ATLAS:2019itm,CMS:2019xai,ATLAS:2018hzj} with up to 139 fb$^{-1}$ of data at 13 TeV. Since no significant excess has been found, limits were set on the coupling and $Z'$ mass based on a simplified model with a Lagrangian
\begin{align}
\cl_{\rm simp}=& -\frac{1}{4} F_{\mu\nu} F^{\mu\nu} - \frac{1}{2} M^2
A_\mu A^\mu -\bar \psi(\frac{\hbar}{i} \gamma^\mu \partial_\mu + m_D)\psi\non
& + A_\mu\bar \psi(g_V +\gamma_ 5 g_A) \gamma^\mu \psi + g_q A_\mu \bar q \gamma^\mu q,
\label{simp}
\end{align}
where $A_\mu$ is the new massive vector boson field with mass $M$ and $F_{\mu\nu}$ is its field strength, $\psi$ is the Dirac fermion with mass $m_D$, $g_V (g_A)$ are the vector (axial vector) coupling of $A_\mu$     
with the Dirac fermion and $g_q$ is the coupling of the vector boson with the SM quarks. The relevant part of the Lagrangian here is the term $g_q A_\mu \bar q \gamma^\mu q$. A comparison between the theoretical and observed cross sections using the simplified model of Eq.~(\ref{simp}) is translated to constraints on $g_q$. We recast the obtained limits to our model parameters as constraints on the kinetic mixing $\delta$ in the case where $Z'$ does not decay to dark fermions, i.e., $m_{Z'}<2m_D$. To do so, we follow the procedure in refs.~\cite{Chang:2022jgo,Bagnaschi:2019djj} and construct the log-likelihood
\begin{equation}
\ln\mathcal{L}=-2\left[\frac{(v_f^{\prime 2}+a_f^{\prime 2})^2\times\text{BR}(Z'\to q\bar q)^2}{g_q^4}\right],   
\label{logll}
\end{equation}
where we have ignored the term in the vertex proportional to $m_f^2/m_{Z'}^2$. The terms in the numerator of Eq.~(\ref{logll}) have a non-trivial dependence on the kinetic mixing and is determined as the one minimizing the log-likelihood function. 

Unlike the simplified model of Eq.~(\ref{simp}), our model is not leptophobic and we have an important decay channel to leptons. So our $Z'$ can have dilepton decays which introduce stringent constraints on the model parameter space. The dilepton channel is cleaner than the dijet since the latter is contaminated by large QCD multijet background. Therefore, the dilepton constraints are much more severe. In constraining the Stueckelberg model, we use the most recent ATLAS dilepton search with Drell-Yan processes~\cite{ATLAS:2019erb}. In recasting limits from dilepton searches, we include possible interference effects with SM processes involving $Z$ and $\gamma$ mediators using a modified version of the code \code{ZPEED}~\cite{Kahlhoefer:2019vhz}.  

For $m_{Z'}\geq 2m_D$, the $Z'\to D\bar D$ channel opens up which means that the branching ratios to leptons and quarks become smaller. So we expect in this case weaker dilepton and dijet limits. However, monojet searches, $pp\to D\bar D+$jet, become relevant, i.e., missing energy recoiling against a hard jet. For this, we use the most recent ATLAS and CMS monojet searches~\cite{ATLAS:2021kxv,CMS:2021snz} to constrain the model parameter space. We updated the monojet module in \code{micrOMEGAs}~\cite{Barducci:2016pcb} with the most recent ATLAS and CMS data and used it as our recasting tool. 

For light $Z'$, searches in the dimuon channel was carried out by CMS~\cite{CMS:2019kiy} as well as LHCb~\cite{LHCb:2017trq,LHCb:2019vmc}, where the latter investigated prompt and long-lived $Z'$. The null results from these experiments are translated into constraints on the kinetic mixing coefficient for a dark photon model, i.e., a $Z'$ which kinetically mixes only with the SM photon. To recast those limits to our model, we implemented the Stueckelberg model in \code{DarkCast}~\cite{Ilten:2018crw,Baruch:2022esd} which has a large repository of the most up-to-date searches on dark photons.

\subsection{BaBar, electron bremsstrahlung and beam dump experiments}

Along with the LHC constraints, many other experiments have investigated a light $Z'$ and set constraints on the kinetic mixing in a simple dark photon model. BaBar analyzed $Z'$ production and decay to visible final states from $e^+e^-$ annihilation, $e^+e^-\to Z'\to e^+e^-~(\mu^+\mu^-)$~\cite{BaBar:2014zli} and to invisible final states~\cite{BaBar:2017tiz}. Furthermore, electron bremsstrahlung experiments such as APEX~\cite{APEX:2011dww} and A1~\cite{Merkel:2014avp} studied the production and decay of $Z'$ to $e^+e^-$ while NA64~\cite{Banerjee:2019pds} studied invisible decays of $Z'$. Beam dump experiments such as E137~\cite{Konaka:1986cb}, E141~\cite{Riordan:1987aw}, E774~\cite{Bjorken:1988as}, KEK~\cite{Bross:1989mp}, and Orsay~\cite{Davier:1989wz} have studied long-lived dark photons. All these limits are part of \code{DarkCast} which we use as our recasting tool in this mass range.

\subsection{Relic density constraints}

Not only is our scan of the parameter space over a wide range of $Z'$ mass but also over a wide range of the kinetic mixing coefficient. This adds a complexity related to the fact that for small $\delta$, the dark sector, comprised of dark fermions and $Z'$, may not be in thermal equilibrium with the SM. This means that one can assume the standard treatment of the freeze-out scenario adopted in codes like \code{micrOMEGAs} and \code{darkSUSY}~\cite{Bringmann:2018lay} only in the case where $\delta$ is large enough to maintain thermal equilibrium. The Boltzmann equation for the number density of $D$ is given by
\begin{equation}
\frac{dn_D}{dt}+3Hn_D=C[f_D],    
\end{equation}
where $C[f_D]$ is the collision term containing DM number-changing processes such as $D\bar D\to f\bar f$ and $D\bar D\to Z'Z'$. For DM lighter than the $Z'$ mediator, the freeze-out DM relic density is set by the annihilation processes into SM fermions. This processes is proportional to $\sim (\delta g_X)^2$. However, for DM heavier than $Z'$, the process $D\bar D\to Z'Z'$, which is proportional to $g_X^4$, becomes kinematically accessible and will set the final DM relic density. In the freeze-out scenario and considering a Maxwell-Boltzmann phase space distribution, the collision term is given by
\begin{equation}
C[f_D]=-\langle\sigma v\rangle(n_D^2-n_{D\,\text{eq}}^2),    
\end{equation}
with the thermally averaged cross section given by
\begin{equation}
\langle\sigma v\rangle=\frac{1}{K_2(x)^2}\int_1^\infty d\tilde{s}\,4x\sqrt{\tilde s}\,(\tilde{s}-1)K_1(2\sqrt{\tilde s}x)\,\sigma_{D\bar D\to XX},    
\end{equation}
where the dimensionless parameters are $x=m_D/T$ and $\tilde s=s/(4m_D^2)$. We use \code{micrOMEGAs}~\cite{Belanger:2014vza,Barducci:2016pcb} to determine the DM relic density in the region where the pure freeze-out mechanism is valid. In our scan, we accept points whose relic density is less than or equal to that measured by the Planck collaboration~\cite{Planck:2018vyg}
\begin{equation}
(\Omega h^2)_{\rm Planck}=0.120\pm 0.001,    
\label{Planck}
\end{equation}
i.e., $f_{\rm DM}=(\Omega h^2)_D/(\Omega h^2)_{\rm Planck}\leq 1$, which keeps the door open for multi-component DM scenarios.

In the very small kinetic mixing regime, the dark species will never reach thermal equilibrium with the SM sector. Despite this, annihilation processes of the type $f\bar f\to D\bar D$ and $f\bar f\to Z'$ can gradually populate the dark sector and set the DM relic density via the freeze-in mechanism~\cite{Hall:2009bx}.  The situation becomes more involved if the coupling $g_X$ among the dark species becomes large enough so that the dark sector reaches thermal equilibrium, i.e., $D$ and $Z'$ enter thermal equilibrium. In this case, the processes $D\bar D\leftrightarrow Z'Z'$ become important and one needs to track the number density of $Z'$ as well. Therefore, the calculation of the relic density now requires solving the coupled Boltzmann equations
\begin{align}
\label{nD}
\frac{dn_D}{dt}+3Hn_D&=-\frac{1}{2}\langle\sigma v\rangle_{D\bar D\to f\bar f}(n_D^2-n_{D\,\text{eq}}^2)-\frac{1}{2}\langle\sigma v\rangle_{D\bar D\to Z'Z'}\left(n_D^2-n_{D\,\text{eq}}^2\frac{n_{Z'}^2}{n_{Z'\,\text{eq}}^2}\right) \nonumber \\
&~~~-\frac{1}{2}\langle\sigma v\rangle_{D\bar D\to Z'}n_D^2+\langle\Gamma\rangle_{Z'\to D\bar D}n_{Z'}, \\
\frac{dn_{Z'}}{dt}+3Hn_{Z'}&=-\langle\sigma v\rangle_{Z'Z'\to D\bar D}\left(n_{Z'}^2-n_{Z'\,\text{eq}}^2\frac{n_D^2}{n_{D\,\text{eq}}^2}\right)+\frac{1}{2}\langle\sigma v\rangle_{f\bar f\to Z'}n_{f\text{eq}}^2 \nonumber \\
&~~~-\langle\Gamma\rangle_{Z'\to f\bar f}n_{Z'}-\langle\Gamma\rangle_{Z'\to D\bar D}\left(n_{Z'}-n_{Z'\,\text{eq}}\frac{n_D^2}{n_{D\,\text{eq}}^2}\right).
\label{nZp}
\end{align}
This treatment is not part of \code{micrOMEGAs} freeze-in routine~\cite{Belanger:2018ccd} and so we use our own numerical calculations with the help of \code{MATLAB} \code{ode15s} to determine the DM relic density in the case when thermal equilibrium cannot be guaranteed. The criteria we use to make this judgment is based on comparing the DM annihilation rate, $n_D\langle\sigma v\rangle_{D\bar D\to f\bar f}$, and/or the $Z'$ decay rate, $\langle\Gamma_{Z'\leftrightarrow f\bar f}\rangle$, to the Hubble parameter $H(T)$. If the rates of both or any of these processes are larger than $H(T)$ then a thermal equilibrium is established between the two sectors and \code{micrOMEGAs}'s freeze-out routine is able to handle this scenario.  

One final comment regarding the validity of Eqs.~(\ref{nD}) and~(\ref{nZp}) is in order. In writing the Boltzmann equations in terms of the number density, one assumes a well-defined phase space distribution (a Maxwell-Boltzmann distribution in this case). This is justified if kinetic equilibrium can be maintained till after chemical decoupling of DM species. In the standard freeze-out scenario, kinetic equilibrium is maintained through efficient elastic scattering of DM with SM particles. However, when the coupling between DM and the visible sector becomes small, elastic scattering may become inefficient causing early kinetic decoupling. In our model, kinetic equilibrium can still hold and this is attributed to the dark sector itself. Once produced, dark matter self-interactions can bring their momentum distribution to a thermal distribution thus allowing one to use the Boltzmann equations for number density (see Fig. 2 in ref.~\cite{Hryczuk:2022gay}). Furthermore, DM-$Z'$ elastic scattering is strong enough (owing to the large gauge coupling $g_X$ in the dark sector) to also keep $Z'$ in kinetic equilibrium. One can then avoid solving the full phase space Boltzmann equations and instead consider the number density equations.

\subsection{Dark matter direct detection}

Experiments on DM direct detection involve the scattering of a DM particle off the nucleus of a heavy material such as xenon. The voluminous experimental apparatus operates for a period of time looking for an excess of nuclear recoil events as a result of DM-nucleon spin-independent (SI) or spin-dependent (SD) scattering. The differential event rate is given by
\begin{equation}
\frac{dR}{dE_R}(E_R)=\frac{\rho_{\rm DM}}{2m_D \mu^2_r}f_{\rm DM}\sigma_{\rm Dn}F^2(E_R)\eta(E_R,t),    
\end{equation}
where $\rho_{\rm DM}$ is the DM density at the Sun's location, $\mu_r$ is the DM-nucleon reduced mass, $F(E_R)$ is the nuclear form factor, $\eta(E_R,t)$ contains all astrophysical information 
\begin{equation}
\eta(E_R,t)=\int_{|\mathbf{v}|>v_{\rm min}}\,d^3\mathbf{v}\frac{1}{v}f(\mathbf{v}),    
\end{equation}
with $f(\mathbf{v})$ the DM velocity distribution and 
\begin{equation}
v_{\rm min}=\sqrt{\frac{m_N E_R}{2\mu^2_r}}.    
\end{equation}
The DM-nucleon cross section, $f_{\rm DM}\sigma_{\rm Dn}$, refers to either the SI or SD cross section. The SD cross section in our model is suppressed compared to the SI one. Furthermore, experimental constraints on SD are much weaker than those on SI and therefore we do not discuss them any further.  We calculate $f_{\rm DM}\sigma_{\rm SI}$ and the total event rate using \code{micrOMEGAs}~\cite{Belanger:2020gnr} which we also use to recast limits from several experiments such as CDMSlite~\cite{SuperCDMS:2015eex}, CRESST-II~\cite{CRESST:2015txj}, CRESST-III~\cite{CRESST:2019jnq}, DarkSide 50~\cite{DarkSide:2018kuk}, LUX 2016~\cite{LUX:2016ggv}, PICO-60~\cite{PICO:2017tgi,PICO:2019vsc}, PandaX~\cite{PandaX-II:2016vec,PandaX-II:2017hlx} and Xenon1T~\cite{XENON:2018voc}. We also take into account the most recent limits from LUX-ZEPLIN (LZ)~\cite{LZ:2022ufs}. Note that there are two diagrams contributing to the SI cross section: the $Z$ and $Z'$ exchange diagrams. Since setting $M_2=0$ would prevent DM from acquiring a millicharge, diagrams with photon exchange are absent as well as the ones with a Higgs since, unlike a Higgs portal model, our DM has no coupling to the Higgs.

\subsection{Dark matter indirect detection}

Even though the DM particles have achieved a constant comoving number density, annihilation of DM particles into the SM can still happen today especially in regions with large density. The annihilation processes can result in $\gamma$ ray emissions, charged particles (such as electrons, positrons or even composite particles such as antiprotons and antideuterons) as well as neutrinos. The detection of charged particles cannot be easily attributed to DM annihilation since it is difficult to trace back the origin of those particles as they are deflected by magnetic fields. However, $\gamma$ ray photons are not affected by magnetic fields and can be an important tool in DM indirect detection. The well known gamma ray excess at the center of our galaxy~\cite{Hooper:2007kb} is an example but it remains debatable as the galaxy center is a rich source of gamma rays from other astrophysical sources.   

As DM annihilate into SM particles, charged final states can radiate off photons which are the source of prompt $\gamma$ ray emission. Photons can also come from the decay of pions after final state quarks have hadronized. The differential photon flux due to DM annihilation for an observation region $d\Omega$ is given by
\begin{equation}
\frac{d\phi}{dE\,d\Omega}(E_\gamma)=\frac{r_{\odot}}{4\pi}\frac{1}{4}\left(\frac{\rho_{\rm DM}}{m_D}\right)^2 J \sum_i f^2_{\rm DM}\langle\sigma v\rangle_i \frac{dN_i}{dE_\gamma},    
\end{equation}
where $r_{\odot}$ is the location of the Sun in the galactic plane, the $J$ factor contains astrophysical information and the photon spectrum $dN_i/dE_\gamma$ due to annihilation to some final state $i$ is determined by \code{PYTHIA}~\cite{Sjostrand:2007gs}. The annihilation cross section as well as the photon flux is calculated using \code{micrOMEGAs} where tabulated results of $dN_i/dE_\gamma$ from \code{PYTHIA} can be found. The results are compared to the 6 years of data from the Fermi-LAT collaboration~\cite{Fermi-LAT:2015att}. Notice that the photon flux is proportional to $f^2_{\rm DM}$ which can help evade those constraints in the case of multicomponent DM, i.e., for $f_{\rm DM}\ll 1$.     

\begin{figure}[H]
\centering
\includegraphics[width=0.495\textwidth]{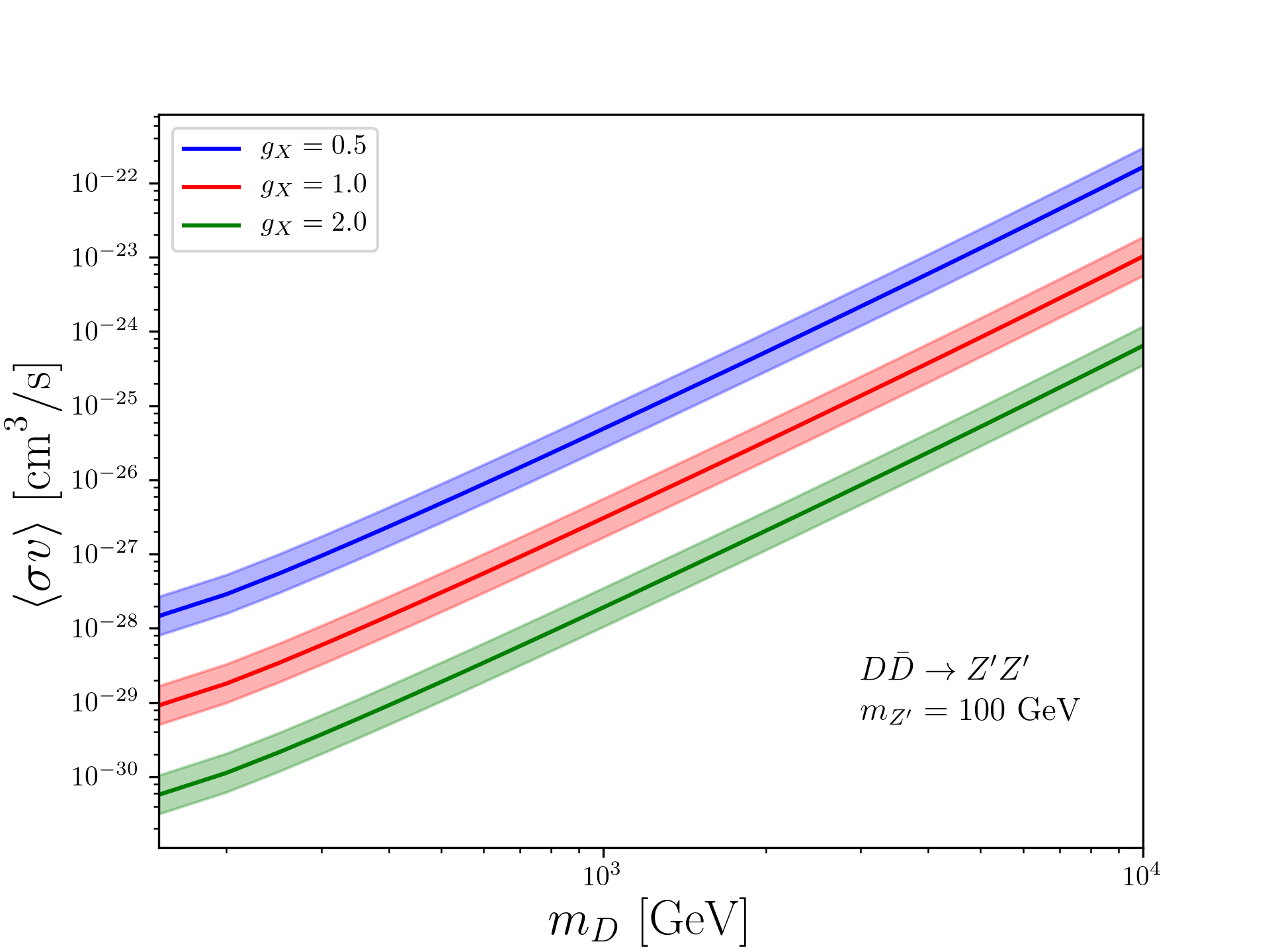}
\includegraphics[width=0.495\textwidth]{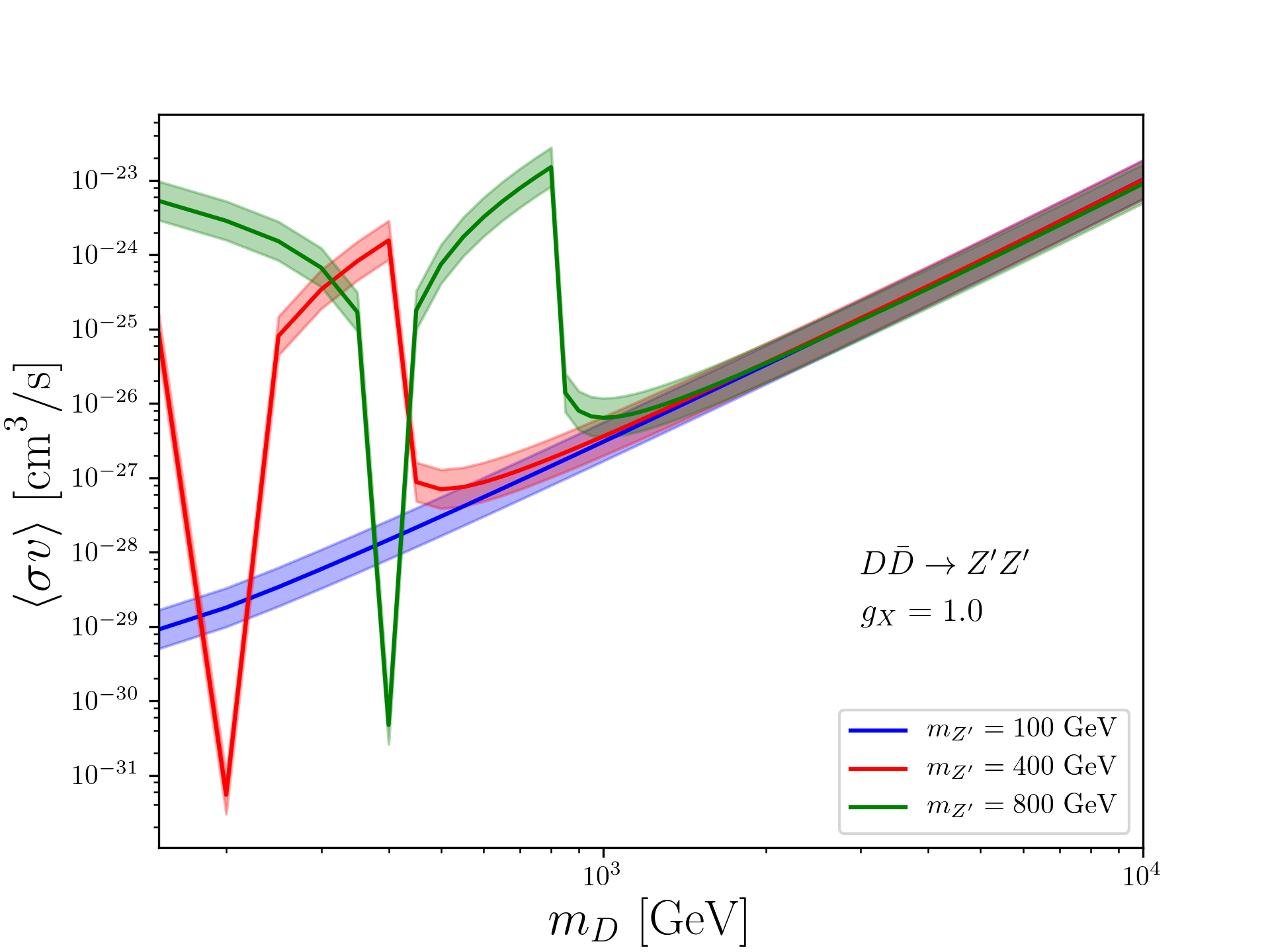}
\caption{The 90\% upper limits on the thermally averaged cross section using the Fermi-LAT data~\cite{Fermi-LAT:2015att} for $Z'Z'$ final state. We show the limits for different $g_X$ values (left panel) and different $m_{Z'}$ (right panel) with the $1\sigma$ uncertainty band.}
\label{fermi-lat}
\end{figure}

To determine the constraints on the DM thermally averaged annihilation cross section from the Fermi-LAT measurements, we use the published data of 15 Milky Way dwarf spheroidal galaxies (dSphs) from the Fermi-LAT collaboration~\cite{Fermi-LAT:2015att}. The published six-year \code{Pass 8} data pertain to the measured photon flux and the bin-by-bin test statistic for each of the dSphs considered in the analysis. Taking the LAT likelihood for target $i$ as $\mathcal{L}_i(\bm{\mu},\bm{\theta}_i|\mathcal{D}_i)$, where $\bm{\mu}$ contains the DM model parameters, $\bm{\theta}_i$ are the nuisance parameters and $\mathcal{D}_i$ are the gamma ray data, we construct the combined likelihood of 15 dSphs as
\begin{equation}
\mathcal{L}(\bm{\mu},\bm{\theta}|\mathcal{D})=\prod_i \mathcal{L}_i(\bm{\mu},\bm{\theta}_i|\mathcal{D}_i). 
\end{equation}
Then we define the test statistic
\begin{equation}
q_s=-2\ln\left[\frac{\mathcal{L}(\bm{\mu_0},\bm{\hat\theta}|\mathcal{D})}{\mathcal{L}(\bm{\hat\mu},\bm{\hat\theta}|\mathcal{D})}\right],    
\end{equation}
which determines the significance of the DM hypothesis. In our definition, $\bm{\mu_0}$ represents the theory parameters under the null hypothesis while the hatted variables are the best fit parameters under the DM hypothesis. The upper limit on $\langle\sigma v\rangle$ is determined for $q_s=2.71$ which represents a 90\% quantile of a $\chi^2$ distribution. We present in Fig.~\ref{fermi-lat} the obtained upper limits for two cases: varying $g_X$ (left panel) and varying $m_{Z'}$ (right panel). For the $D\bar{D}\to Z'Z'$ channel, we notice a strong dependence on $g_X$ and milder dependence on $m_{Z'}$.

The right panel of Fig.~\ref{fermi-lat} shows two sharp dips in the upper limits for $m_{Z'}=400$ GeV and 800 GeV. Notice the dips occur at $m_D=m_{Z'}/2$ which corresponds to the resonance region for the process $D\bar{D}\to f\bar{f}$. In this case, the annihilation cross section becomes very large which results in a large photon flux. Therefore, the analysis of Fermi-LAT data produces very stringent bounds in this region and so the upper limit on $\langle\sigma v\rangle$ becomes very small, i.e., more stringent as can be seen from the dips. The limits we present in Fig.~\ref{fermi-lat} will be translated into constraints on the kinetic mixing and dark photon mass.

\section{Exclusion limits}\label{sec:limits}

In this Section we give the exclusion limits on the Stueckelberg $Z'$ model from the various experimental constraints discussed in the previous section. The results pertain to the heavy and light $Z'$ for a wide range of kinetic mixing. Before we discuss the results, note that the $m_{Z'}$ in our model is not a free parameter as it depends on the kinetic mixing $\delta$ and the mass parameter $M_1$. In the case of vanishing mass mixing, recall that the $Z'$ mass is $m_{Z'}^2=(q\pm p)/2$, where~\cite{Feldman:2007wj}
\begin{align}
\label{p-eq}
p&=\sqrt{\left[M_1^2 c^2_\delta+\frac{v^2}{4}\left(g_Y^2 c^2_\delta+g_2^2\right)\right]^2-M_1^2 v^2 c^2_\delta(g_Y^2+g_2^2)}, \\
q&=M_1^2 c^2_\delta+\frac{v^2}{4}\left(g_Y^2 c^2_\delta+g_2^2\right).
\label{q-eq}
\end{align}
The positive sign in $m^2_{Z'}$ corresponds to a heavy $Z'$ while the negative is for a light $Z'$. 

\subsection{Heavy $Z'$ bosons}

\begin{figure}[H]
\centering
\includegraphics[width=0.495\textwidth]{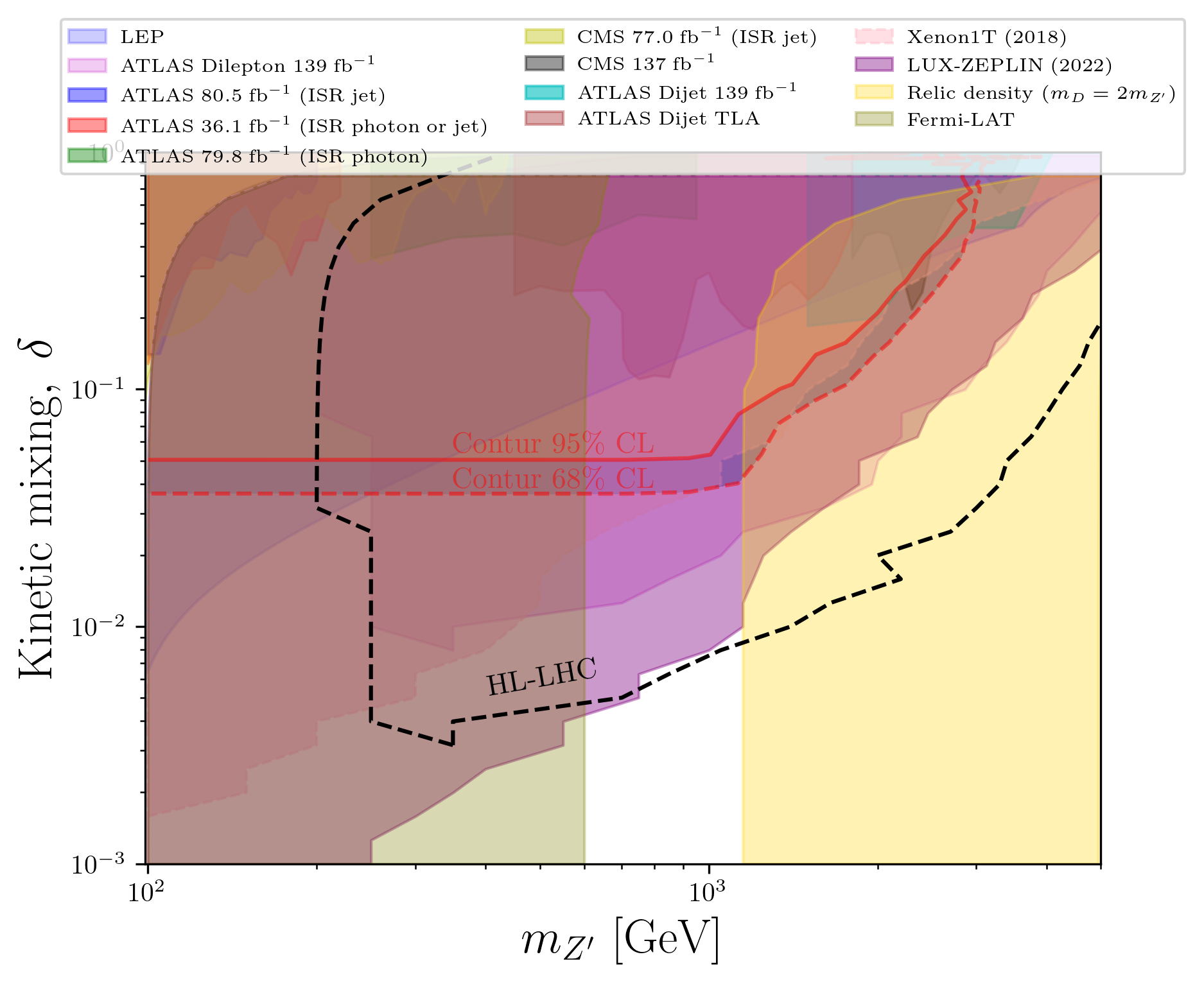}
\includegraphics[width=0.495\textwidth]{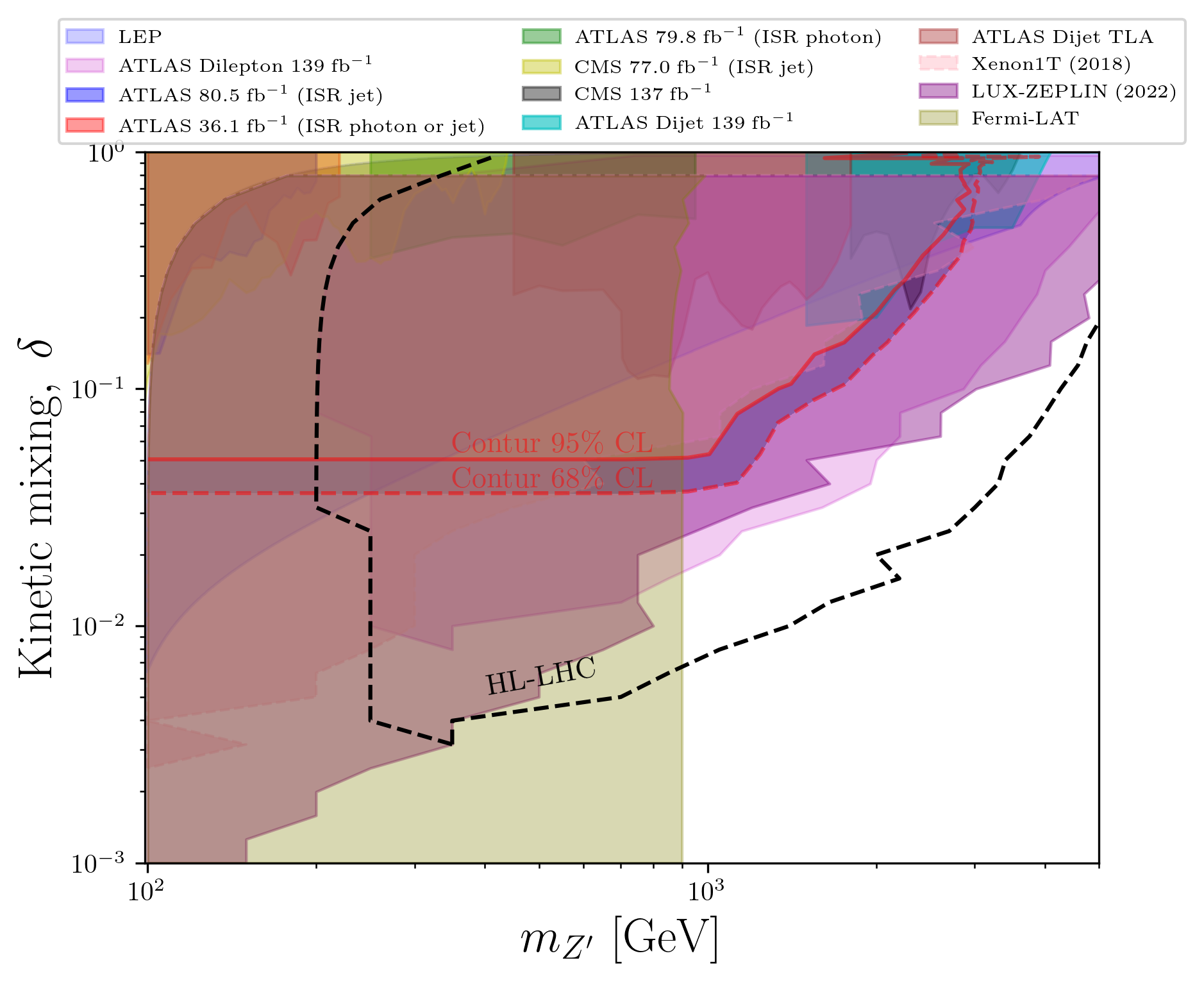}
\caption{Exclusion limits from LHC analyses, LEP, DM relic density, direct and indirect detection experiments in the kinetic mixing-$Z'$ mass plane for $m_D=2m_{Z'}$ (left panel) and $m_D=1.5m_{Z'}$ (right panel). The dashed black contour shows the projected reach of HL-LHC in the dilepton channel. Note that the region bordered by a red solid (dashed) line represents the 95\% (68\%) CL from \code{Contur}.}
\label{heavy-MZp-no-decay}
\end{figure}

We first consider the case of a heavy $Z'$ with $m_D>2m_{Z'}$, whose decay width to SM fermions is given by
\begin{equation}
\Gamma_{Z'\rightarrow f\bar{f}}=\frac{N_c\, g_2^2}{48\pi\,\cos^2\theta}m_{Z'}\sqrt{1-\frac{4m_f^2}{m_{Z'}^2}}\left[v_f^{\prime 2}+a_f^{\prime 2}+\frac{2m_f^2}{m_{Z'}^2}(v_f^{\prime 2}-2a_f^{\prime 2})\right],
\label{eq:Zff}
\end{equation}
where $N_c=3$ for quarks, 1 for leptons and $1/2$ for neutrinos. 
In Fig.~\ref{heavy-MZp-no-decay} we show the relevant constraints in the kinetic mixing-$Z'$ mass plane which include dijet and dilepton searches from ATLAS and CMS as well as the LEP constraint. We also show constraints from direct detection experiments (Xenon1T and LZ), Fermi-LAT and the DM relic density. The latter only appears in the left panel for reasons we discuss thereafter. The figure also exhibits the 95\% and 68\% CL regions obtained from \code{Contur} using precision measurements of SM processes. We identify in Fig.~\ref{heavy-M1-no-decay} of Appendix~\ref{app:C} the different analyses pools giving the highest sensitivity for exclusion. In this figure we plot $M_1$ rather than $m_{Z'}$.   

In plotting Fig.~\ref{heavy-MZp-no-decay}, we chose $m_D=2m_{Z'}$ (left panel) and $m_D=1.5m_{Z'}$ (right panel) which means that $Z'$ has no dark decays and the process $D\bar D\to Z'Z'$ is kinematically allowed. As a result, the DM relic density is set by the latter annihilation process and so we expect the relic density to have mild to no dependence on the kinetic mixing $\delta$. In fact, for a fixed $m_{Z'}$, the relic density is solely determined by $g_X$ which we take here to be 1.0 in the left panel. The gold-colored area in the left panel of Fig.~\ref{heavy-MZp-no-decay} shows the region excluded by the relic density. One can see that for a fixed $m_{Z'}$ changing $\delta$ has no effect on the relic density. But this begins to change for $\delta\gtrsim 0.3$ as the boundary starts curving rightward for larger $m_{Z'}$. The reason is that $m_{Z'}$ is not an independent parameter. According to Eqs.~(\ref{p-eq}) and~(\ref{q-eq}), $m_{Z'}$ depends on the scanning parameters $M_1$ and $\delta$. For small $\delta$, $m_{Z'}\approx M_1$, but for larger values, $m_{Z'}\gg M_1$. In this case, the kinetic mixing starts affecting the relic density because it causes the $Z'$ mass to change. Larger $m_{Z'}$ means that the process $D\bar D\to Z'Z'$ becomes less efficient and so DM does not readily depletes causing the relic density to shoot up. The relic density constraint disappears from the right panel for the choice $m_D=1.5m_{Z'}$ and $g_X=2.0$ since in this case all points have a relic density smaller than Eq.~(\ref{Planck}). As one can clearly see from Fig.~\ref{heavy-MZp-no-decay}, DM (in)direct detection and dilepton searches as well as the DM relic density (for the left panel) are the most constraining limits on the parameter space. However, there remains parts of the parameter space that can still be explored as seen from the right panel. We draw the projected reach in the kinetic mixing-mass plane at HL-LHC in the dilepton channel (black dashed curve). We will explore in section~\ref{sec:lhc} the discovery potential of HL-LHC as a validity of this region drawn here. One more comment is in order regarding the \code{Contur} limits from precision measurements of SM processes. As seen from Fig.~\ref{heavy-MZp-no-decay}, those limits are very competitive and are more stringent than the dijet limits and come close to the reach of the dilepton limits near 1 TeV.

\begin{figure}[H]
\centering
\includegraphics[width=0.495\textwidth]{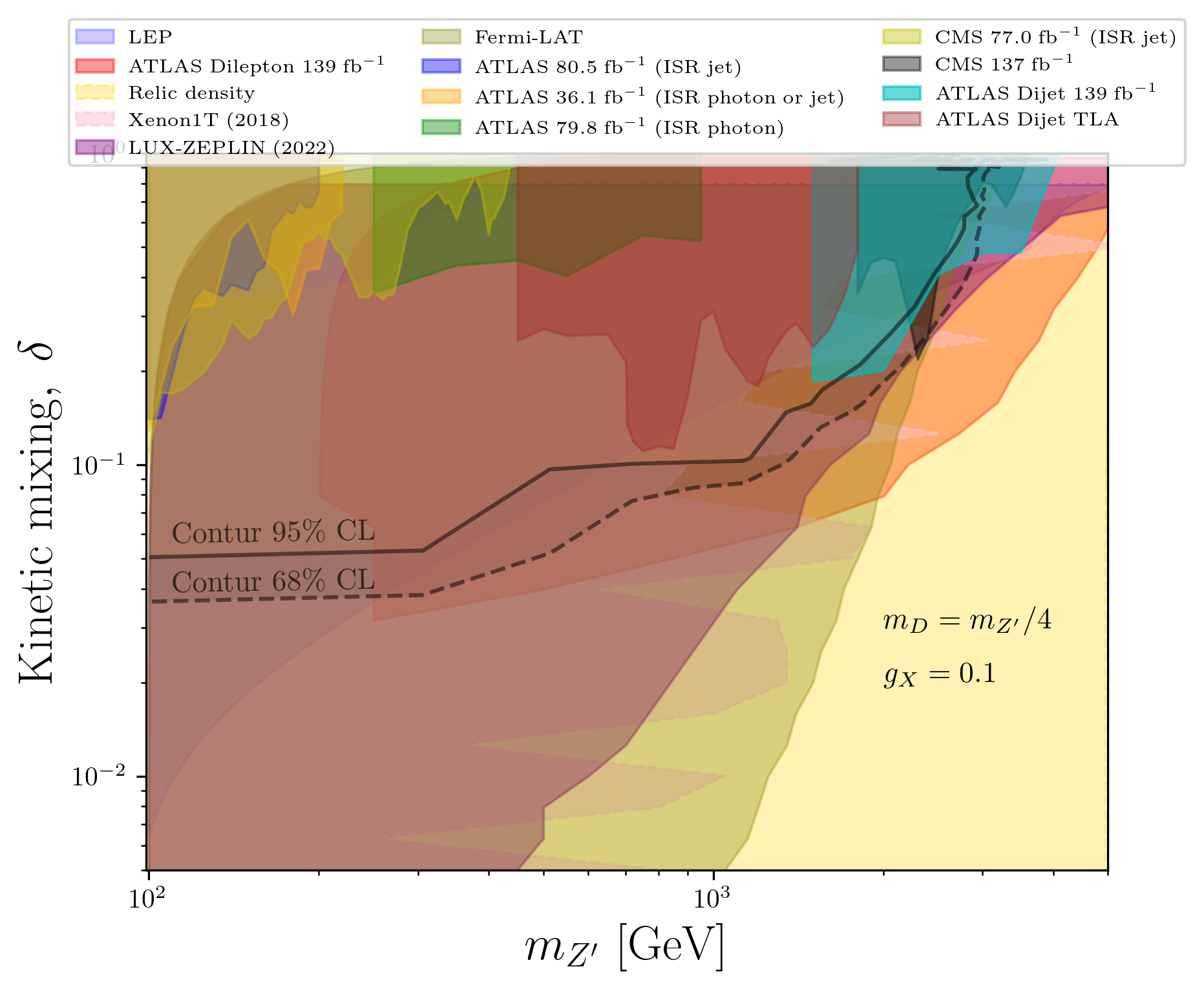}
\includegraphics[width=0.495\textwidth]{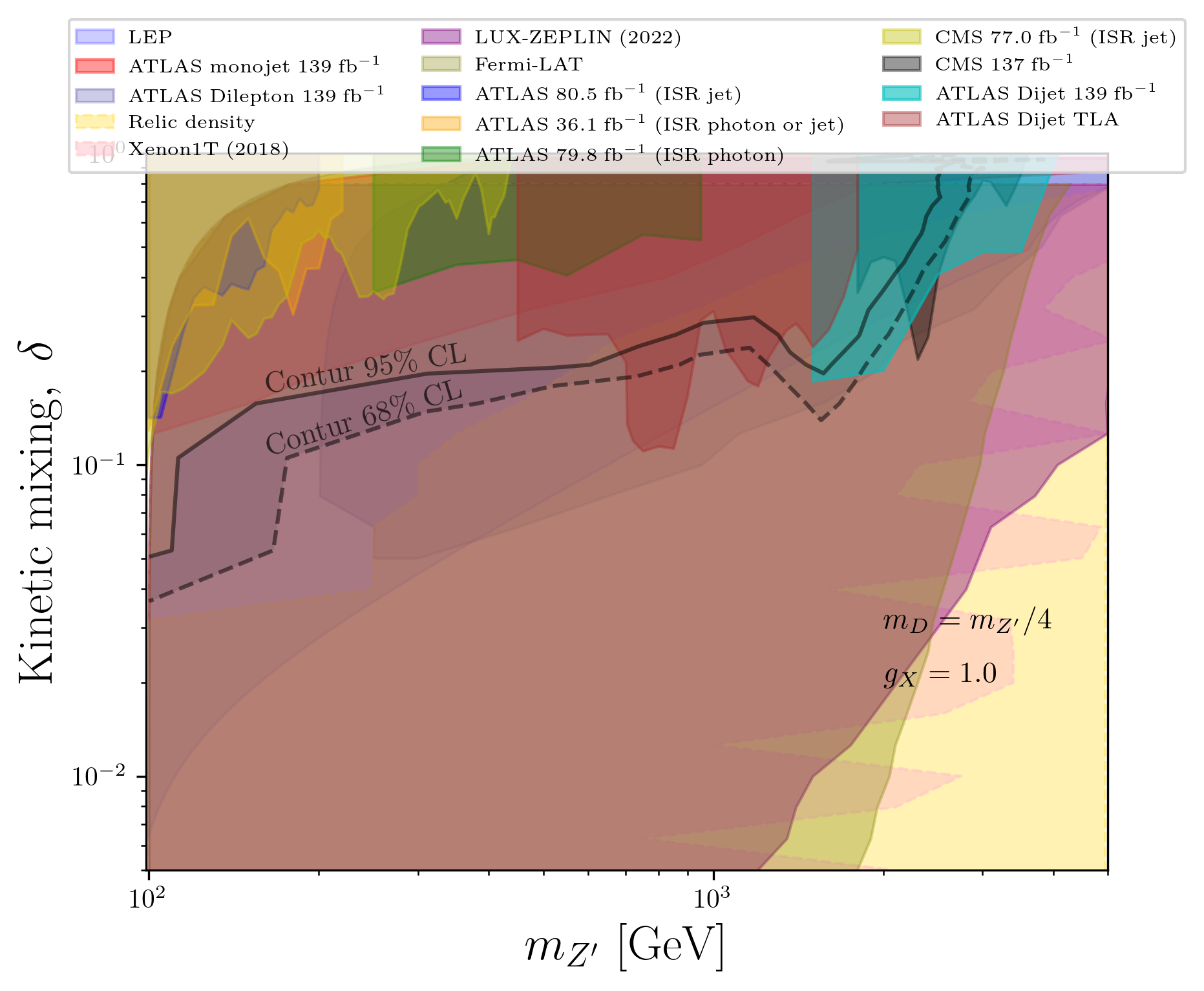} 
\caption{Limits from DM direct and indirect detection experiments, relic density and LHC constraints from ATLAS dilepton, dijet and monojet searches at 139 fb$^{-1}$ for $m_D=m_{Z'}/4$ and $g_X=0.1$ (left panel) and $g_X=1.0$ (right panel). The 95\% CL and 68\% CL contours from \code{Contur} and LEP constraint are also shown.}
\label{heavy-MZp-w-decay}
\end{figure}

We now allow $Z'$ to decay to the dark fermions by taking $m_D=m_{Z'}/4$, thus opening a new decay channel with a decay width
\begin{equation}
\Gamma_{Z'\rightarrow D\bar{D}}=\frac{m_{Z'}}{12\pi}g_X^2(\mathcal{R}_{11}-s_\delta \mathcal{R}_{21})^2\sqrt{1-\frac{4m_D^2}{m_{Z'}^2}}\left(1+\frac{2m_D^2}{m_{Z'}^2}\right).
\label{eq:ZDD}
\end{equation}
This will weaken the limits from dilepton and dijet searches as one can clearly see from Fig.~\ref{heavy-MZp-w-decay}, with the left panel corresponding to $g_X=0.1$ and the right one to $g_X=1.0$. The LEP constraint as well as limits from precision calculations of SM processes obtained from \code{Contur} are added along with limits from direct and indirect detection experiments. Assuming thermal production of DM, the final DM relic density for this setup is set by annihilation to SM fermions since now $D\bar D\to Z'Z'$ is not accessible. The processes $D\bar D\to f\bar f$ depend on $\delta^2 g_X^2$ and so the relic density can only be satisfied for large enough $\delta$ as shown in Fig.~\ref{heavy-MZp-w-decay}. This region, however, is already excluded by LEP and dilepton searches. Even for larger $g_X$, the yellow region opens up but is still not enough to evade LHC constraints. One can thus see that this parameter space is completely ruled out. But this only corresponds to the DM-$Z'$ mass relation used here. So it is important to check other values by scanning over $m_D$ and $m_{Z'}$ instead.    

\begin{figure}[H]
\centering
\includegraphics[width=0.495\textwidth]{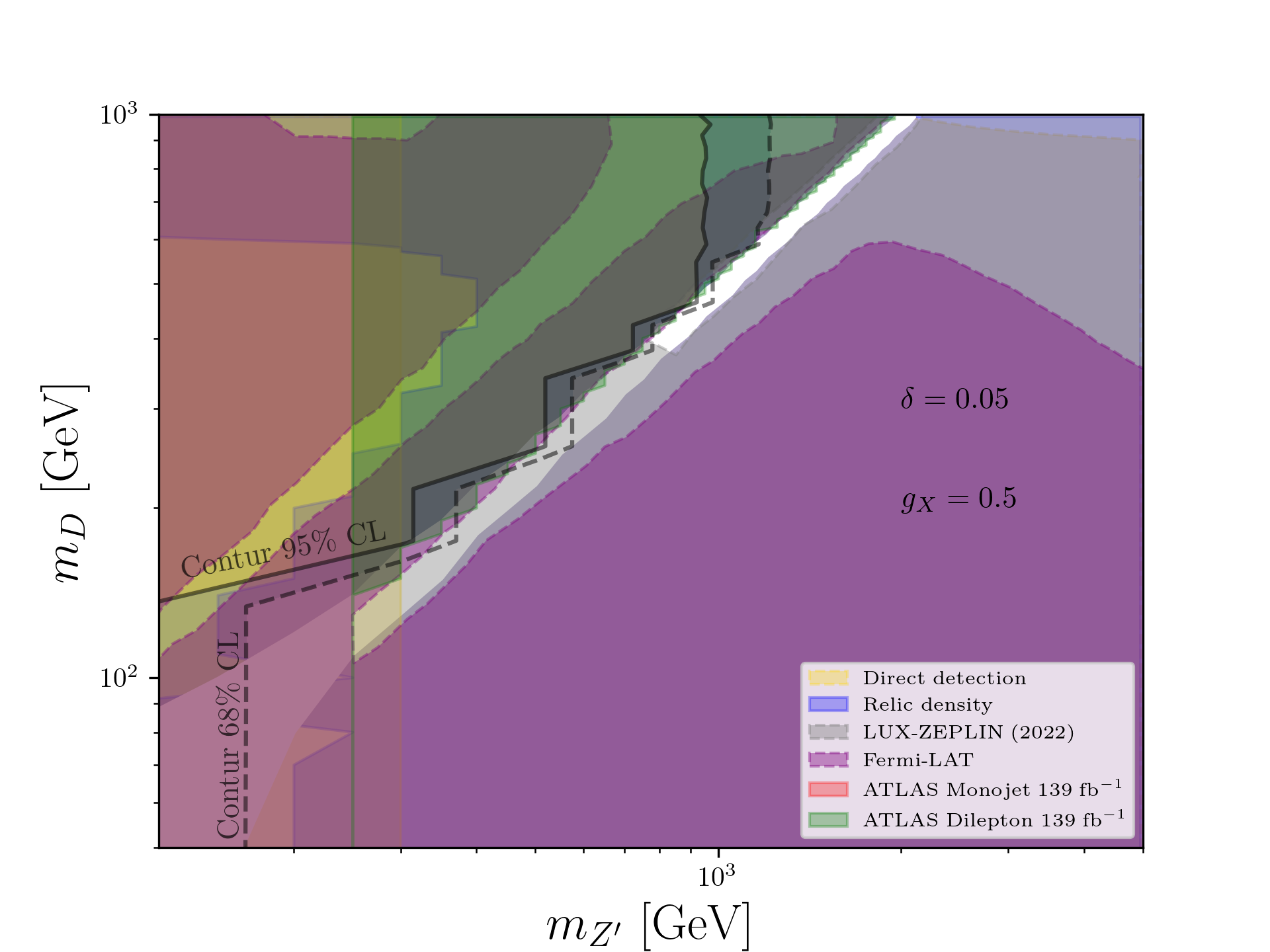}
\includegraphics[width=0.495\textwidth]{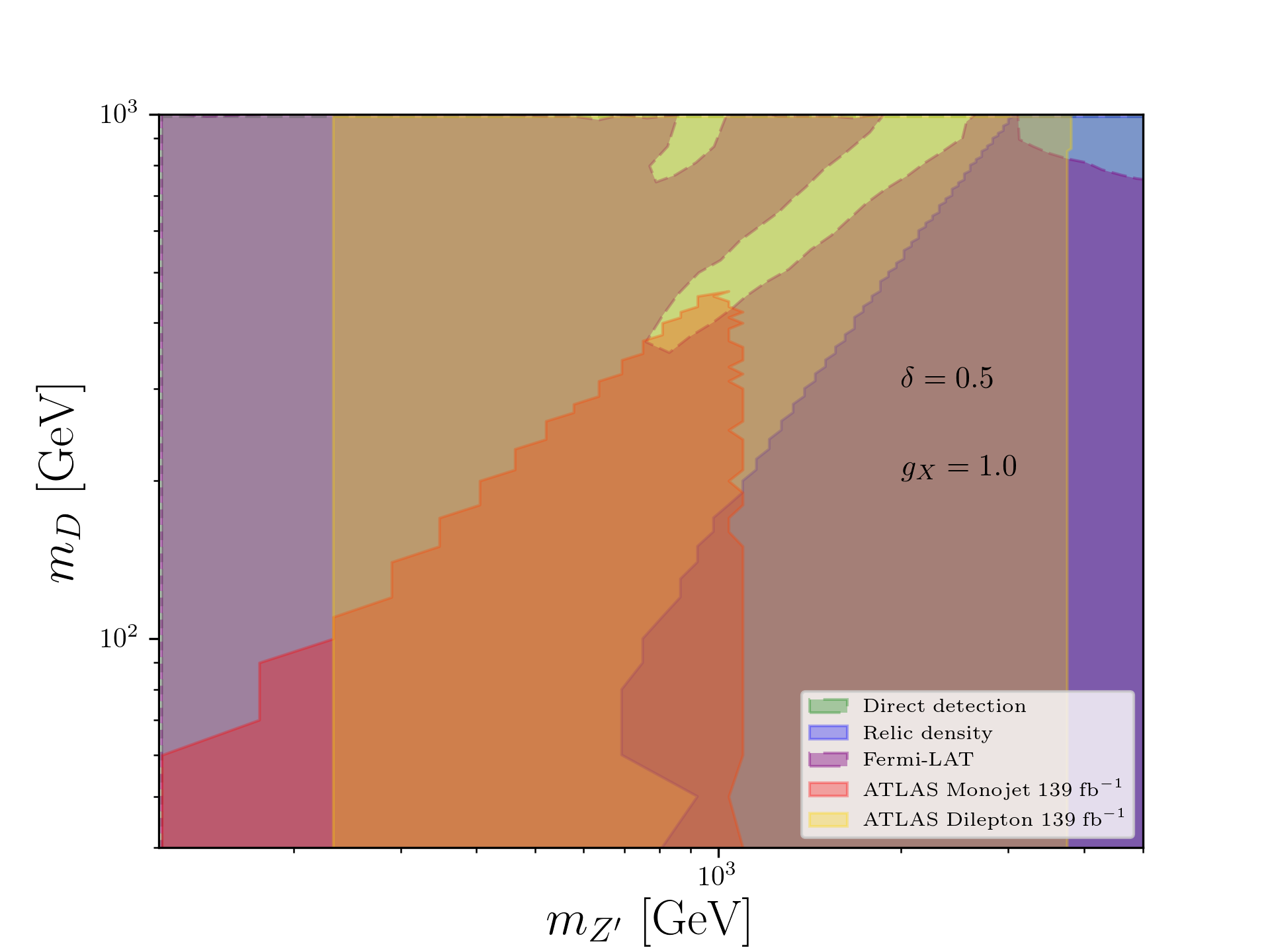}
\caption{Constraints on the Stueckelberg model in the DM-$Z'$ mass plane for two choices of $\delta$ and $g_X$. Limits from direct detection experiments, monojet and dilepton searches as well as the relic density constraint eliminate most of the parameter space except near the resonance region for smaller $\delta$ and $g_X$ (left panel).}
\label{DM-Zp-plane}
\end{figure}

Fig.~\ref{DM-Zp-plane} shows the different experimental constraints in the DM-$Z'$ mass plane for a specific choice of $\delta$ and $g_X$. In addition to the previous limits, monojet searches become relevant here and are also included. In the left panel of Fig.~\ref{DM-Zp-plane} one can see that most of the parameter space is excluded except for a narrow region along the diagonal which corresponds to the resonant decay $Z'\to D\bar D$. The relic density constraint misses this region because the annihilation channel $D\bar D\to f\bar f$ proceeds on-shell which efficiently depletes the DM density. In the right panel of Fig.~\ref{DM-Zp-plane} we exhibit the constraints for higher values of $\delta$ and $g_X$. In this case the monojet and dilepton constraints become very strong that they exclude all the parameter space including the resonance region. 

The important takeaway here is that the Stueckelberg parameter space is severely constrained for the case of dark decays of $Z'$. This is due to DM-related constraints, i.e., the relic density as well as direct and indirect detection experiments. The only available window is near the resonance region for smaller couplings. For heavier DM mass, no $Z'$ dark decays occur and the process $D\bar D\to Z'Z'$ become available which constitutes an important channel for depletion of the DM abundance. For a large choice of $g_X$, the relic density constraint can be severely weakened but it does not come with out a cost. The Fermi-LAT constraint becomes stronger but would still allow a considerable part of the parameter space to still be open. One can see here that the combination of all these constraints points to a thermalized dark sector, i.e., a sector in which its constituent species are in thermal equilibrium. In other words, the DM relic density is controlled by annihilation within the dark sector itself rather than annihilation into SM fermions.

\subsection{Light $Z'$ bosons}

\begin{figure}[H]
\centering
\includegraphics[width=0.65\textwidth]{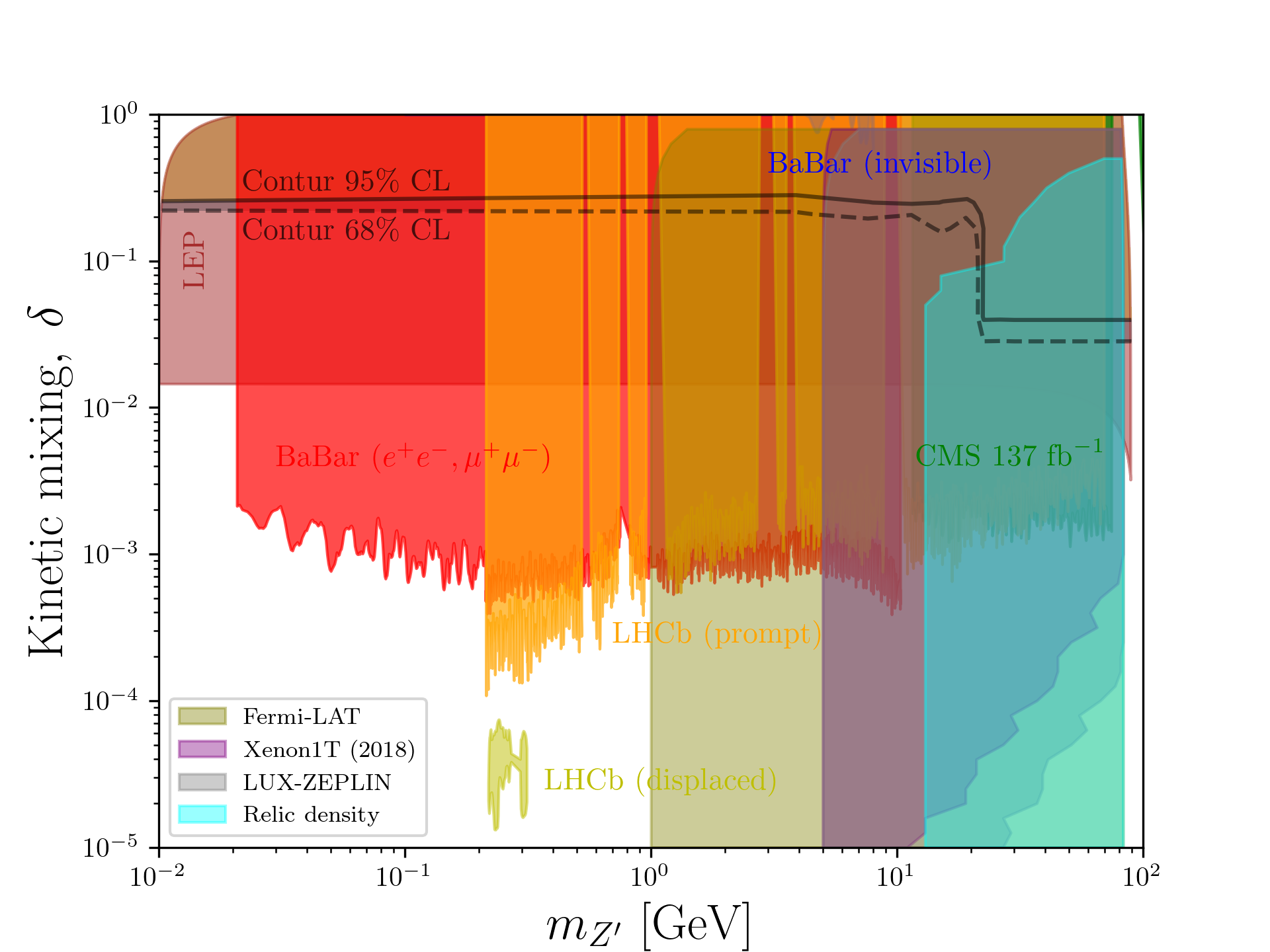}
\caption{The different recasted experimental constraints from LEP, BaBar, CMS, LHCb and (in)direct detection experiments for the light $Z'$ case. We also show the 95\% CL (solid line) and the 68\% CL (dashed line) excluded regions from \code{Contur} as well as the relic density constraint. Here $m_D=2m_{Z'}$ and so the limits correspond to the case of $Z'$ decaying to SM fermions only and no dark decays of the $Z'$.}
\label{light-MZp}
\end{figure}

In this Section we present the experimental limits on our model for the light $Z'$ case, i.e., $m_{Z'}<m_Z$. For this mass range, limits from BaBar and LHCb become important. Since most of the limits are in the $\delta<0.01$ region, one can safely take $M_1\approx m_{Z'}$ so that the vector and axial-vector couplings reduce to the simple forms
\begin{align}
v_f^{\prime}&\simeq \frac{\sin\theta_W}{1-(m_{Z'}/m_Z)^2}\left[2Q_f (\sin^2\theta_W-1)+\frac{m_{Z'}^2}{m_Z^2}(2Q_f-T_{3f})\right]\delta, \\
a_f^{\prime}&\simeq \frac{m_{Z'}^2}{m_{Z'}^2-m_{Z}^2}\sin\theta_W T_{3f}\delta.
\end{align}
Also in the small $\delta$ limit, $(\mathcal{R}_{11}-s_\delta \mathcal{R}_{21})\sim 1$ so that $g_{Z'}= g_X Q_X (\mathcal{R}_{11}- s_{\delta} \mathcal{R}_{21})\sim g_X Q_X$, where we take $Q_X=1$. This approximation is implemented in \code{DarkCast} to derive the experimental upper limits for our model in the kinetic mixing-mass plane as shown in Fig.~\ref{light-MZp}.  

\begin{figure}[H]
\centering
\includegraphics[width=0.495\textwidth]{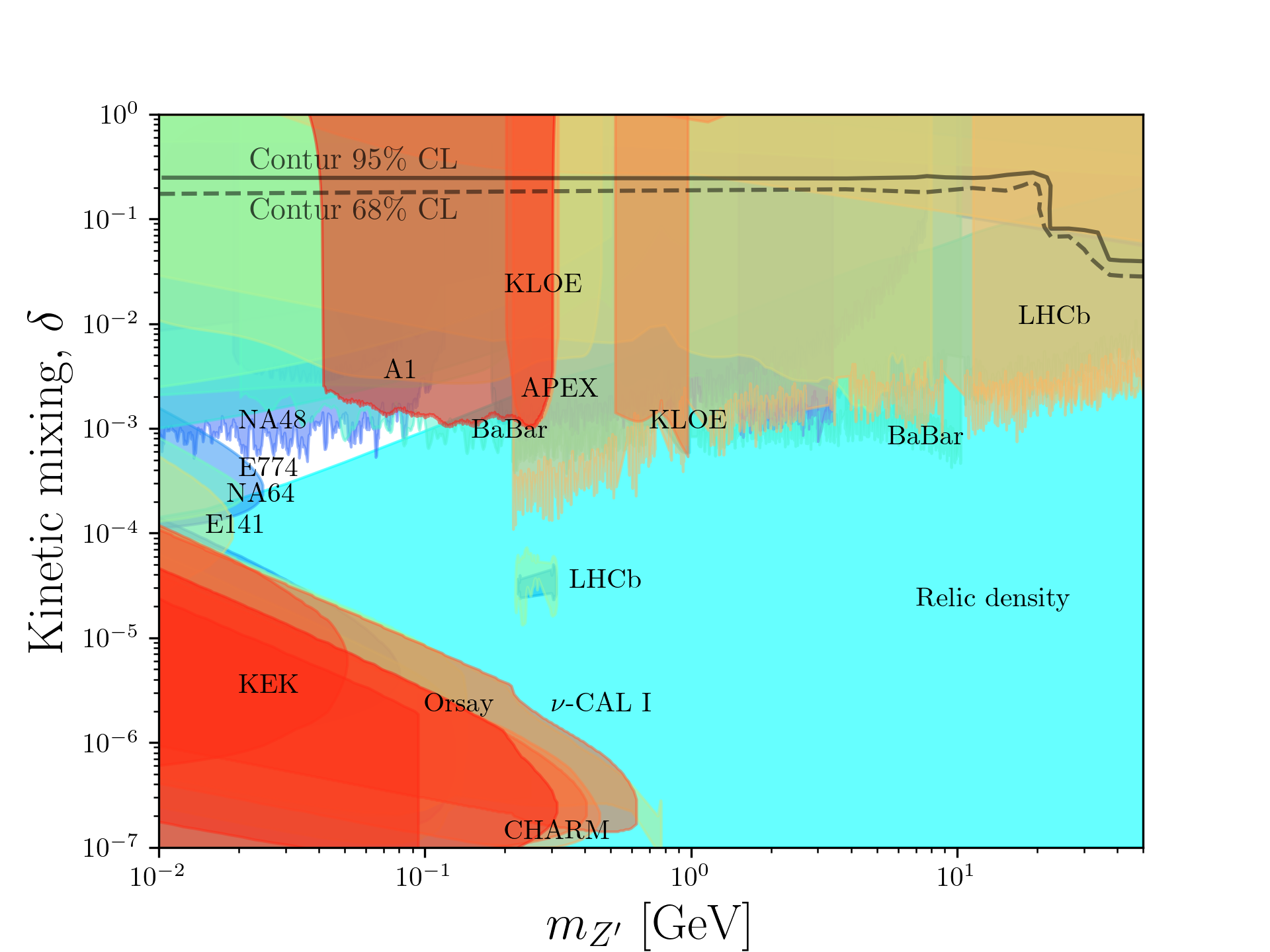}
\includegraphics[width=0.495\textwidth]{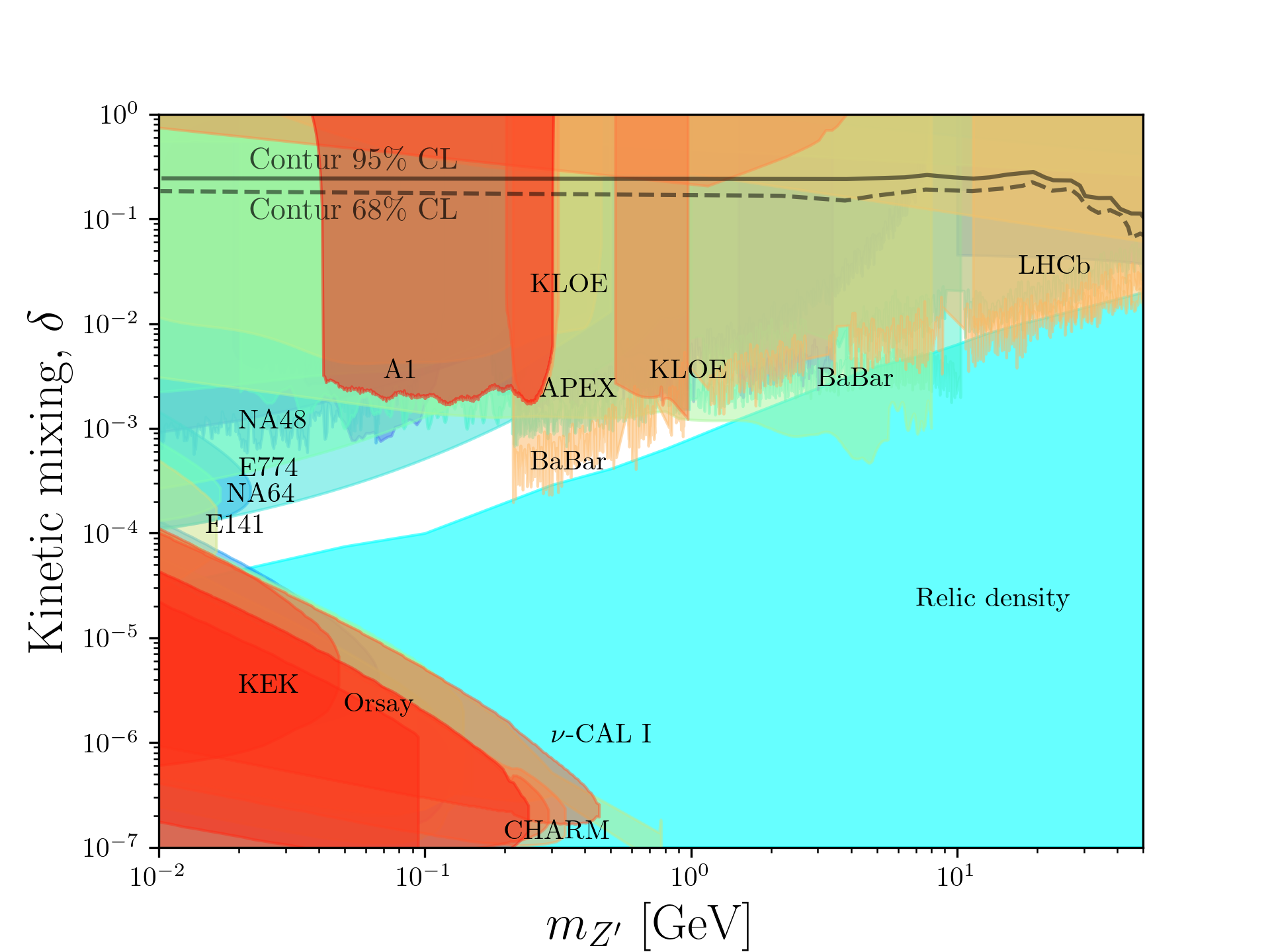}
\caption{Constraints from various visible and invisible limits for a Stueckelberg dark photon in the mass range less than the $Z$ boson mass. The left panel corresponds to $g_X=0.1$ and the right for $g_X=1.0$ and with the common assumption that $m_D=m_{Z'}/4$. The light blue region corresponds to the relic density constraint..}
\label{light-MZp-decay}
\end{figure}

One can see in Fig.~\ref{light-MZp} that for $m_{Z'}>1$ GeV the parameter space is excluded assuming that $g_X$ is large enough to produce a thermal dark sector. Again here the channel $D\bar{D}\to Z'Z'$ is responsible for setting the DM relic density. Now we allow the dark photon to decay to DM by setting $m_D=m_{Z'}/4$. In this case, DM annihilation via $D\bar{D}\to f\bar{f}$ becomes important and since we are considering small values of $\delta$ as shown in Fig.~\ref{light-MZp-decay}, thermal equilibrium between the dark sector and the visible sector cannot be guaranteed. One can easily check this by comparing $n_D\langle\sigma v\rangle_{D\bar D\to f\bar f}$ to the Hubble parameter $H$, where $n_D\langle\sigma v\rangle_{D\bar D\to f\bar f}<H$ for all temperatures means the two sectors have not reached thermal equilibrium\footnote{Note that one should also compare $n_f^{\rm eq}\langle\sigma v\rangle_{f\bar f\to Z'}$ to $H(T)$.}. To determine the DM relic density, we numerically solve the coupled Boltzmann equations, Eqs.~(\ref{nD}) and~(\ref{nZp}), assuming the freeze-in mechanism. In other words, owing to the small value of the kinetic mixing, we assume that DM has a negligible initial abundance in the early universe and that this abundance gradually increases due to annihilation of SM particles, i.e., $f\bar{f}\to D\bar{D}$ and $f\bar{f}\to Z'$. In Fig.~\ref{light-MZp-decay}, we show in light blue the relic density constraint for two values of $g_X$ along with the numerous limits from other experiments. We notice that the available parameter space grows with increasing $g_X$ due to increased DM depletion. 

\begin{figure}[H]
\centering
\includegraphics[width=0.32\linewidth]{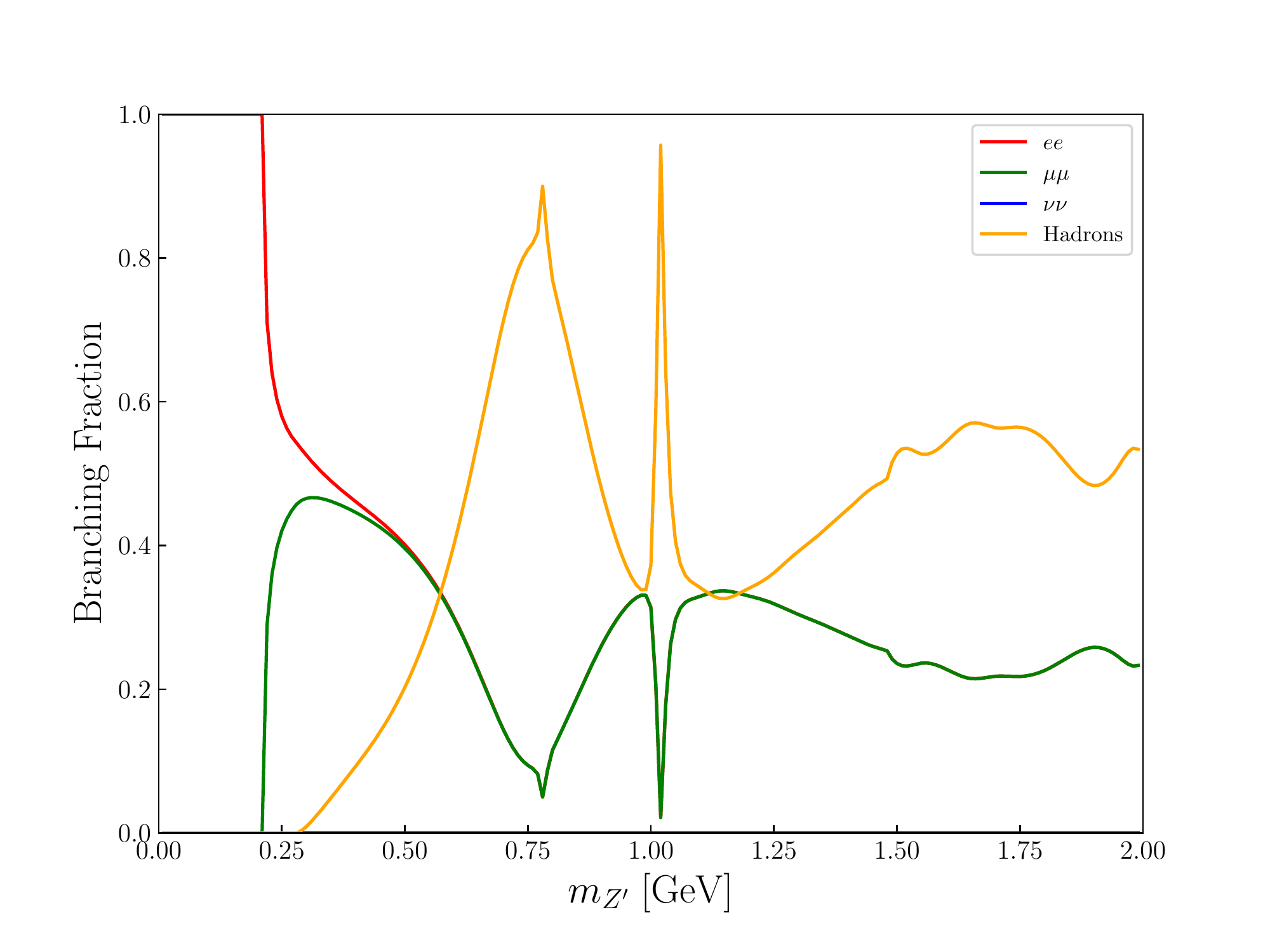}
\includegraphics[width=0.32\linewidth]{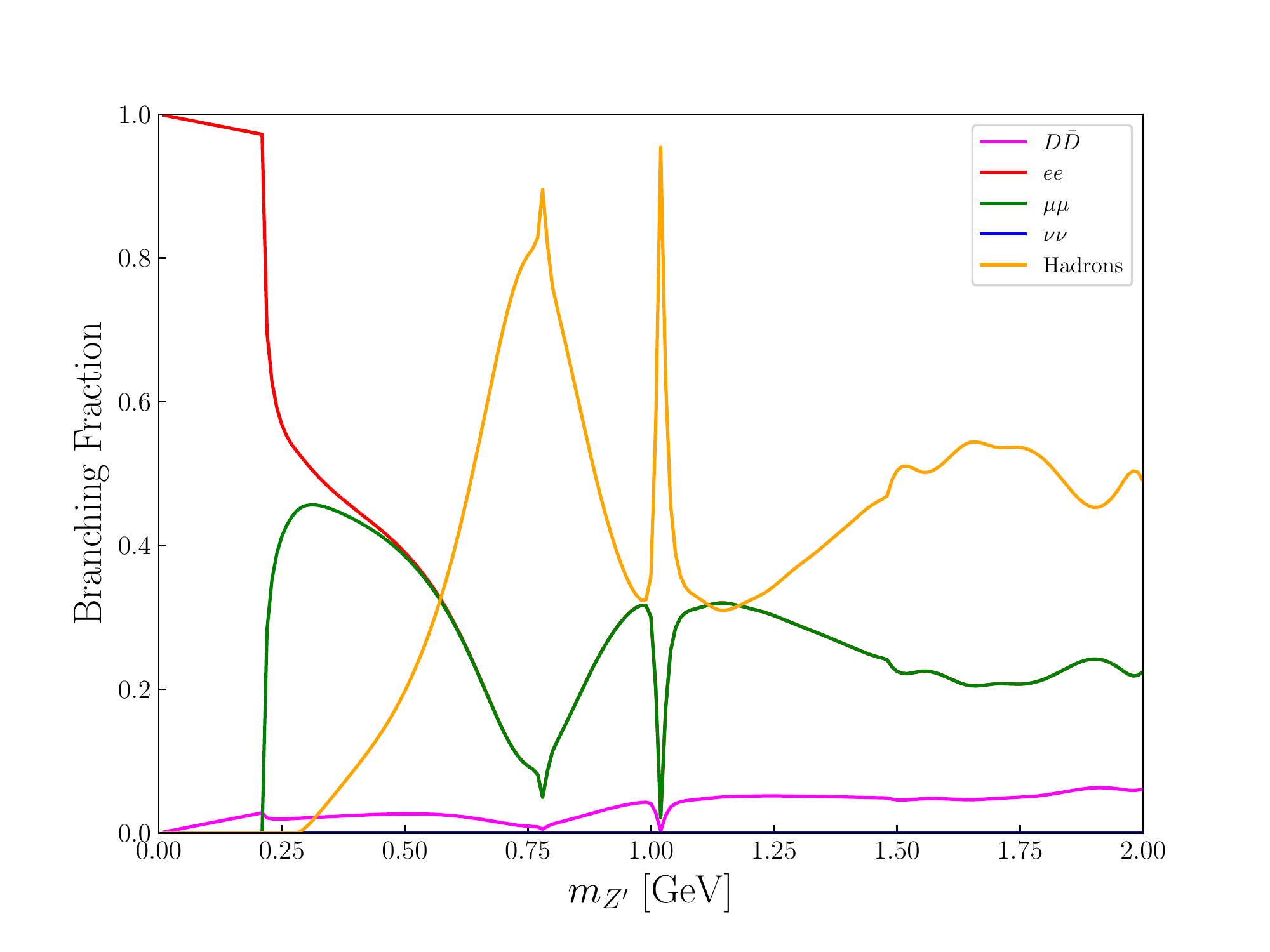}
\includegraphics[width=0.32\linewidth]{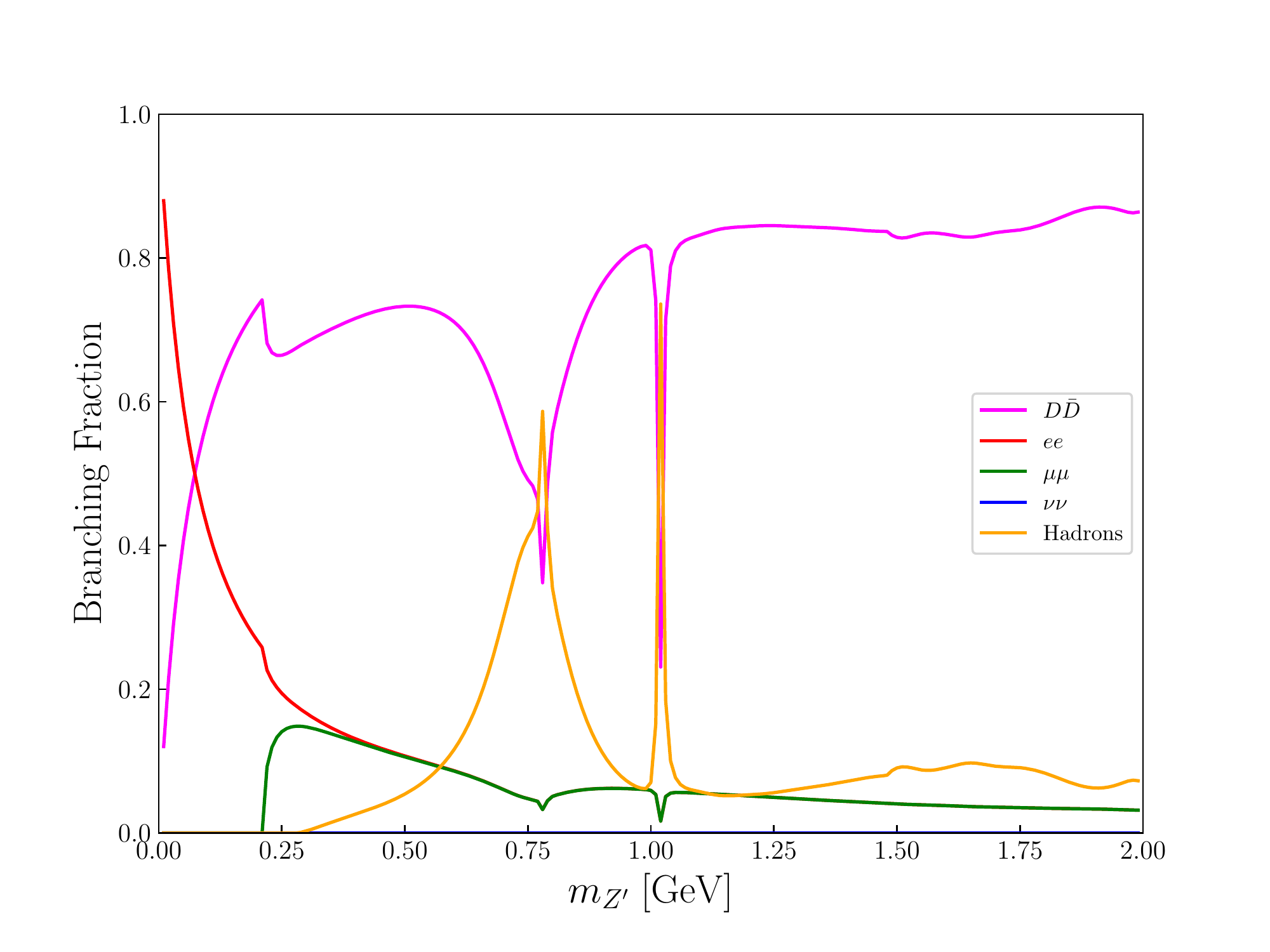}
\caption{Branching fractions of the Stueckelberg dark photon calculated using \code{DarkCast}. In the figure legends, $\nu\nu$ represents the sum of all three generations of neutrinos, `Hadrons' represents the sum of all the branching ratios of the hadronic channels. The left panel shows the branching fractions when there's no dark decays while the middle and right panels show the case of dark decays for $g_X=0.1$ and $g_X=1.0$, respectively. Since we take $m_D=m_{Z'}/4$, the hidden sector decays
are present for any $Z'$ mass. The analysis shows the strong effect of $g_X$  on the branching ratio in the hidden sector from minimal at $g_X=0.1$ to dominant at $g_X=1.0$. }
\label{fig:br}
\end{figure}

Note how the limits change as we increase the value of $g_X$ going from the left panel to the right. Since the main experiments look into the decays of $Z'$ (visible and invisible decays), it is important to properly model such decays especially that in this mass region, hadronic decays of the dark photon can no longer be simply the sum of the decays to $q\bar{q}$. Here we have used \code{DarkCast} to accurately determine the dark photon branching ratios. This is shown in Fig.~\ref{fig:br} for the cases of no dark decays (left panel) and with dark decays for $g_X=0.1$ (middle panel) and $g_X=1.0$ (right panel). 

Overall, we arrive here at the same conclusion we drew in the heavy $Z'$ case. For a kinetically mixed dark photon with dark decays, the parameter space is severely constrained while more available parameter space remains for the case of DM heavier than the mediator $Z'$. 

\section{Discovery potential of the Stueckelberg $Z'$ boson at the HL-LHC}\label{sec:lhc}

The $Z'$ mass reach projected at HL-LHC and exhibited in Fig.~\ref{heavy-MZp-no-decay} shows that a 1 TeV $Z'$ with $\delta\sim\mathcal{O}(10^{-2})$ and heavier can be probed. In this section, we perform a detailed analysis for potential discovery of a Stueckelberg $Z'$ with a TeV scale mass at HL-LHC. Here we focus on the dilepton channel which, despite having a lower cross section than the dijet channel, is clean and one can reconstruct the dilepton invariant mass with much less SM background. However, the challenge here is the naturally small production cross section due to a TeV mass $Z'$ and a small kinetic mixing.  

\begin{figure}[H]
\centering
\includegraphics[width=0.65\textwidth]{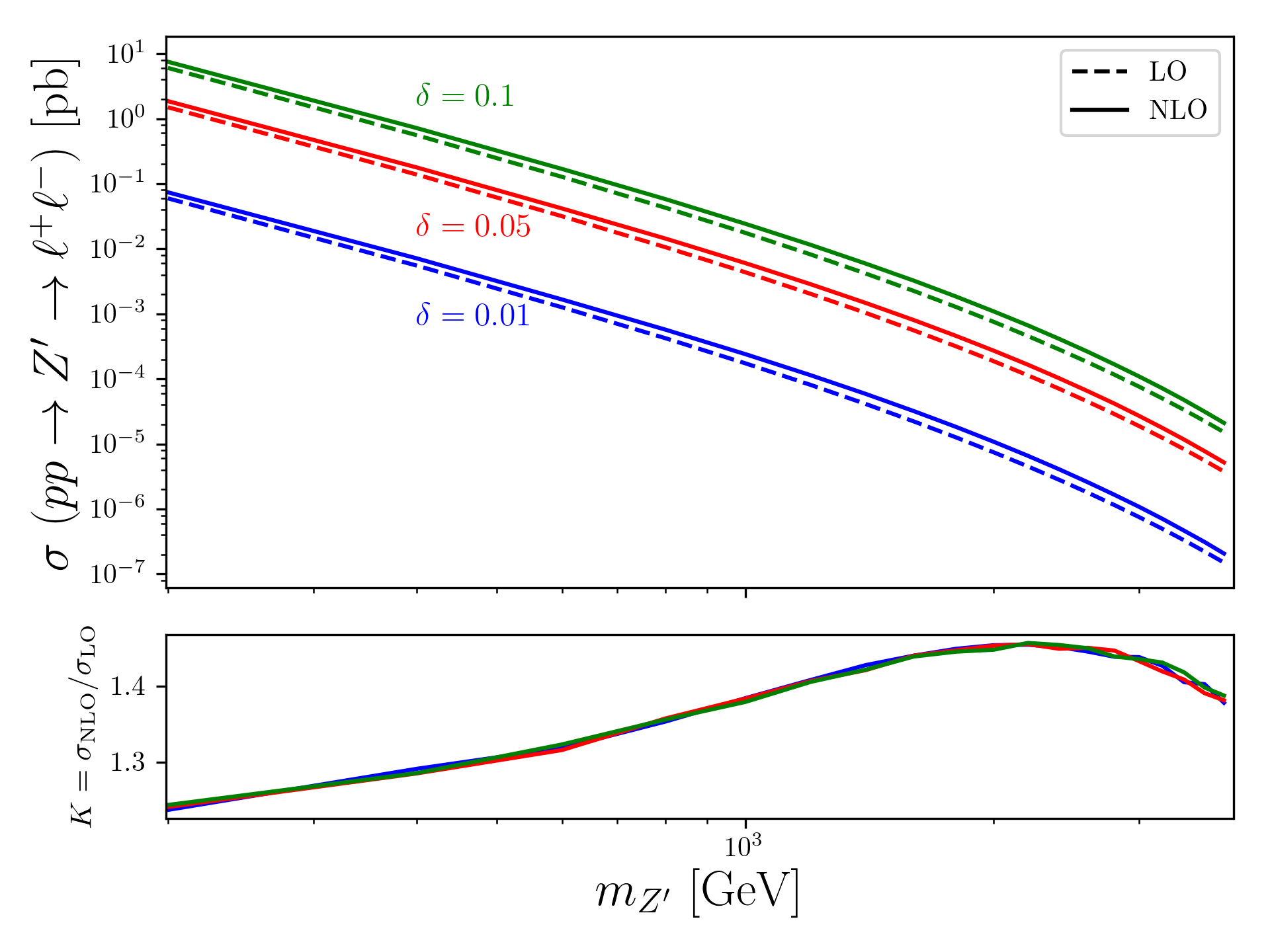}
\caption{The LHC production cross section of a Stueckelberg $Z'$ as a function of $m_{Z'}$ in the dilepton channel at $\sqrt{s}=14$ TeV. The cross section is calculated at LO and NLO for three values of the kinetic mixing $\delta$. The $K$ factor is shown in the bottom plot.}
\label{fig:xs}
\end{figure}

The production cross section of a Stueckelberg $Z'$ is evaluated at NLO. To do so, we implement the model in \code{FeynRules}~\cite{Alloul:2013bka} interfaced with \code{NLOCT}~\cite{Degrande:2014vpa} and \code{FeynArts}~\cite{Hahn:2000kx}. The obtained \code{UFO} files are used in \code{MadGraph5\_aMC@NLO} to determine the LO and NLO cross section of $pp\to Z'\to \ell^+\ell^-$ (see ref.~\cite{Fuks:2017vtl} for a model-independent analysis of $W'$ and $Z'$ production at the LHC). The LO and NLO cross section of the process as a function of $m_{Z'}$ is shown in Fig.~\ref{fig:xs} for three values of $\delta$. The $K$ factor defined as $K=\sigma_{\rm NLO}/\sigma_{\rm LO}$ is shown in the bottom panel. We notice a factor of $\sim 1.5$ increase from the LO prediction for $m_{Z'}\sim 2$ TeV.  

The signal Monte Carlo events are generated at LO using \code{MadGraph5} and the cross section is scaled accordingly using the obtained $K$ factors. The dilepton final state has several sources of SM backgrounds. The dominant ones are: diboson (mainly $WW$), $Z/\gamma^*$+jets, dilepton from off-shell vector boson decay, $t\bar{t}$, single top and top associated production with a vector boson. Background and signal events are generated with \code{MadGraph5} and showered with \code{PYTHIA8}~\cite{Sjostrand:2014zea,Bierlich:2022pfr} (adding ISR and FSR jets and ignoring multiparticle interactions). Detector effects are included using \code{Delphes}~\cite{deFavereau:2013fsa} which implements \code{FastJet}~\cite{Cacciari:2011ma} for jet clustering with the anti-$k_t$~\cite{Cacciari:2008gp} algorithm and jet radius $R=0.4$.    

The event preselection is based on a recent analysis by the ATLAS collaboration~\cite{ATLAS:2019erb}. Electrons with a transverse energy of $E_T>30$ GeV and located within $|\eta|<2.47$ are selected, while muons with a transverse momentum $p_T>30$ GeV and $|\eta|<2.5$ are kept. Events are required to contain at least two same flavor leptons. Candidate events with two muons are required to have oppositely charged muons while such a requirement is not forced on electrons because high $E_T$ electrons suffer from a higher probability of charge misidentification. If an event contains more than two leptons, then the electrons (muons) with the highest $E_T$ ($p_T$) are kept. If an event is found to contain two lepton pairs, then the electron pair is retained because the ATLAS detector has a better resolution and higher efficiency for electrons. Next, the dilepton invariant mass is reconstructed and a minimum cut of 220 GeV is applied as means to reject the overwhelming SM events near the $Z$ pole mass.   

\subsection{Cut-and-count analysis}

\begin{figure}[H]
\centering
\includegraphics[width=0.495\textwidth]{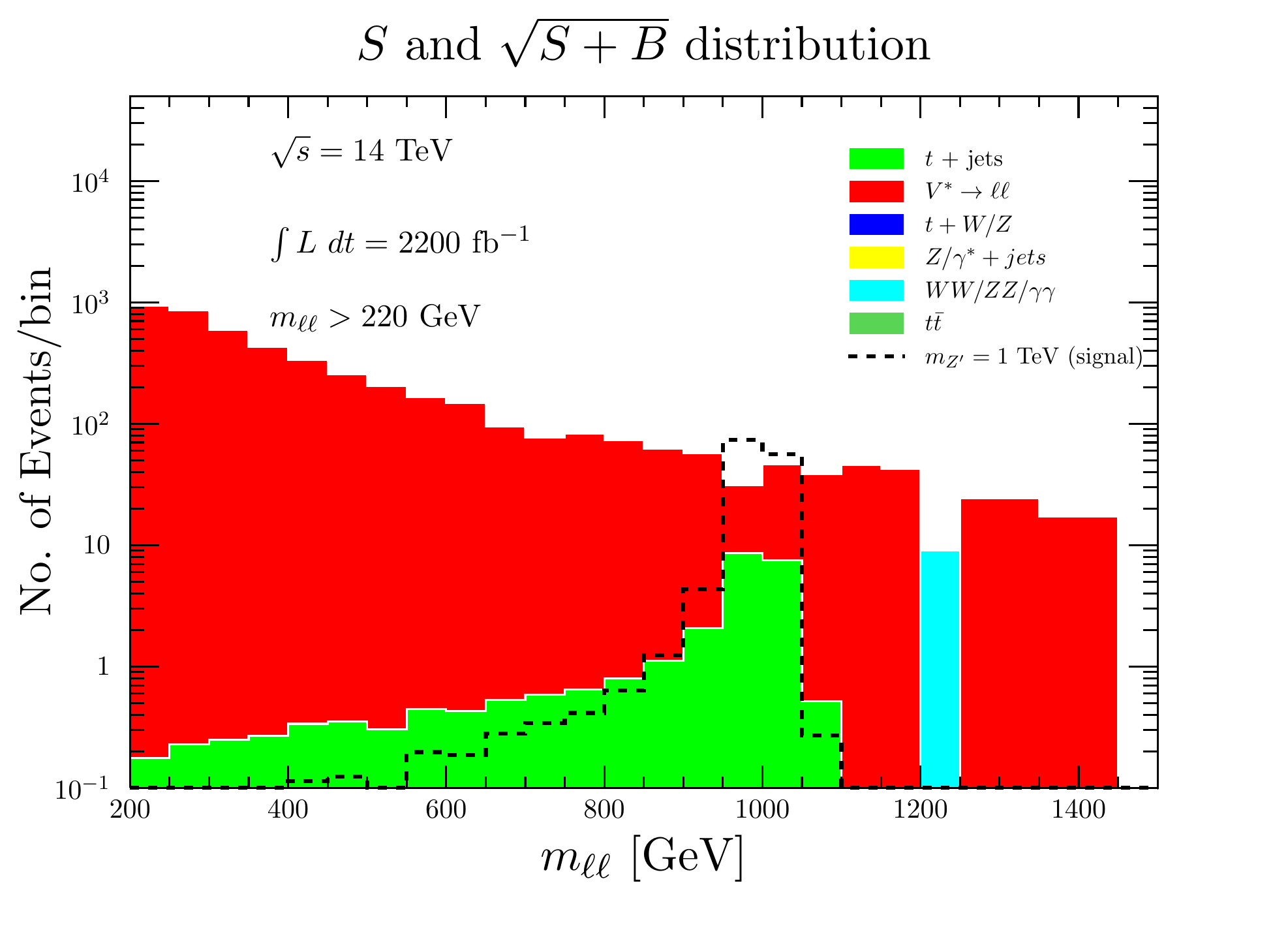}
\includegraphics[width=0.495\textwidth]{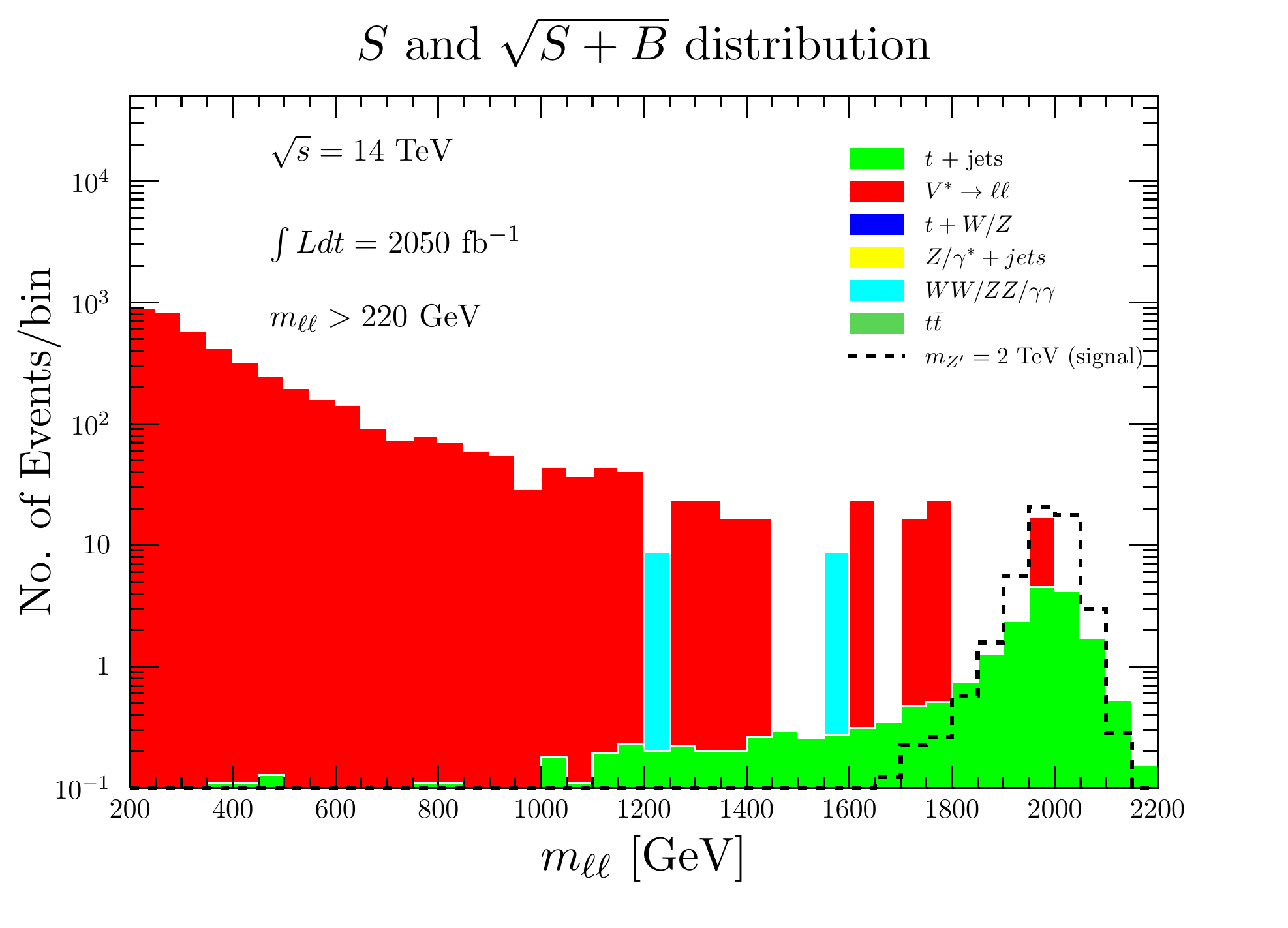}
\caption{The dilepton invariant mass distribution for the signal and background samples. The analysis is for the case of two benchmarks discussed in the text, i.e., for $(m_{Z'},\delta)$ cases $(1$ TeV, $10^{-2}$) and $(2$ TeV, $3\times 10^{-2})$ using the cut-and-count method with cuts on the variables of Eq.~(\ref{cutcount}). This method is found not efficient for discovery with a projected luminosity of 3000 fb$^{-1}$ at HL-LHC. A better technique is discussed in the next section.}
\label{fig:mll}
\end{figure}

We select two benchmarks $(m_{Z'},\delta)$ which lie within the region of reach for HL-LHC. The benchmarks $(1$ TeV$, 10^{-2})$ and $(2$ TeV$, 3\times 10^{-2})$ have NLO cross sections of 0.241 fb and 0.097 fb, respectively. For the cut-and-count analysis, we employ the kinematic variables
\begin{equation}
m_{\ell\ell},~~~E_{T1},~~~E_{T2},~~~p_{T1},~~~p_{T2},
\label{cutcount}
\end{equation} 
where the subscripts `1' and `2' indicate leading and subleading leptons, respectively. We perform a cut-and-count analysis where different cuts based on the above kinematic variables are implemented with the aim to maximize the $S/\sqrt{S+B}$ figure of merit. Using 3000 fb$^{-1}$ as the maximum integrated luminosity projected at HL-LHC, the figure of merit never reaches the $5\sigma$ limit required for discovery. The main culprit here is the irreducible SM dilepton background from off-shell decays of a vector boson. This is clear from Fig.~\ref{fig:mll} where we show the signal and background distribution in the invariant dilepton mass. 

\subsection{Boosted decision tree analysis}

To have a better discrimination between the SM background and the signal, we use a boosted decision tree (BDT) which is part of the \code{TMVA} (Toolkit for Multivariate Analysis)~\cite{Hocker:2007ht,Speckmayer:2010zz} framework embedded in \code{ROOT}~\cite{Antcheva:2009zz,Antcheva:2011zz}. We train a BDT on the signal and background events using the above kinematic variables. The training phase is following by a testing phase carried out on statistically independent Monte Carlo samples of the signal and background events, where the algorithm determines a new kinematic variable called the `BDT response'. This variable is a powerful discriminant necessary to enhance $S/\sqrt{S+B}$. To include the effect of uncertainties, we actually use
\begin{equation}
\frac{S}{\sqrt{S+B+(\delta_S S)^2+(\delta_B B)^2}}
\label{merit}
\end{equation}
as the figure of merit for a $5\sigma$ discovery. Here, $\delta_S$ ($\delta_B$) represents the systematic uncertainty in the signal (background) which we take to be 10\% (20\%). 

\begin{figure}[H]
\centering
\includegraphics[width=0.495\textwidth]{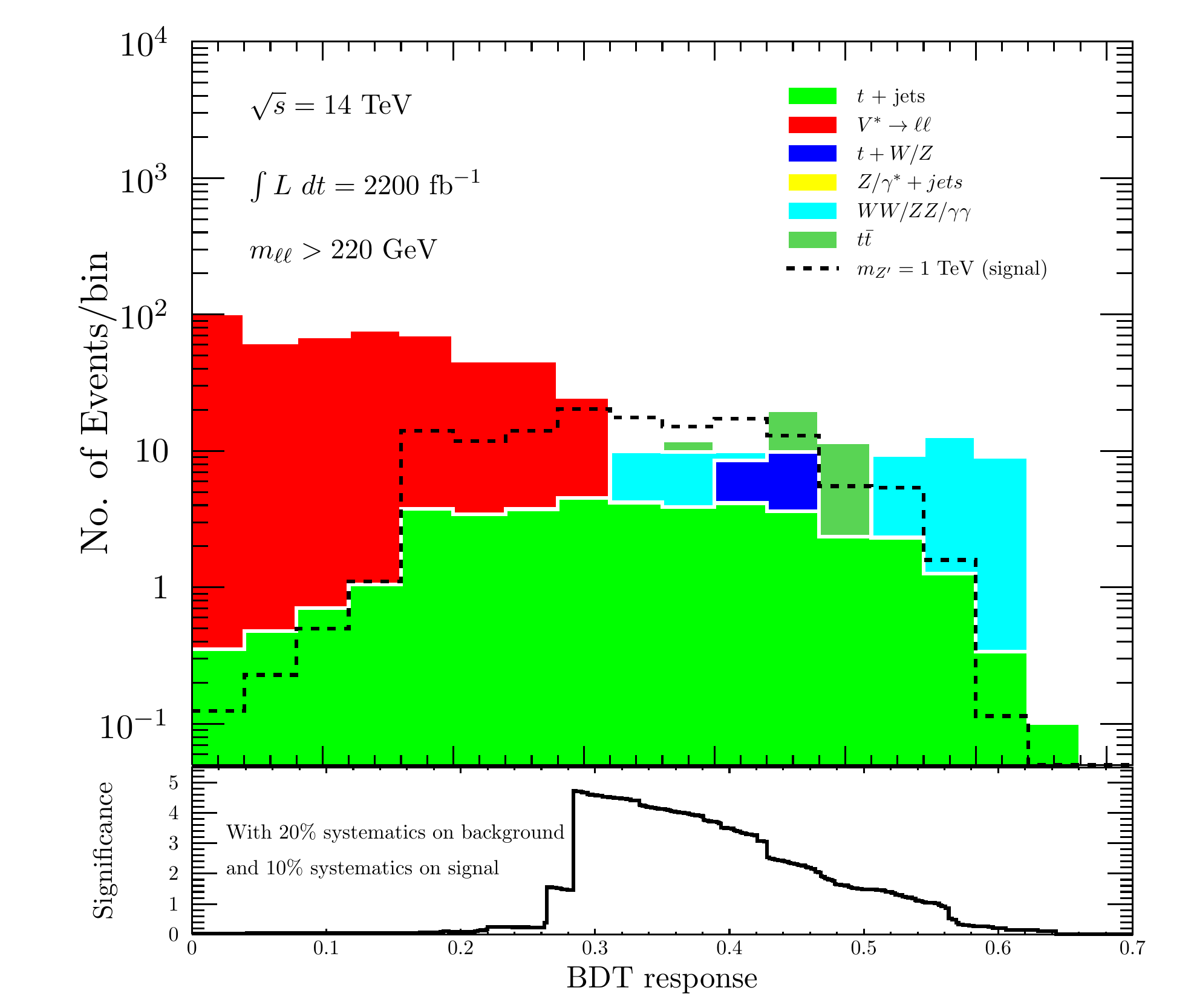}
\includegraphics[width=0.495\textwidth]{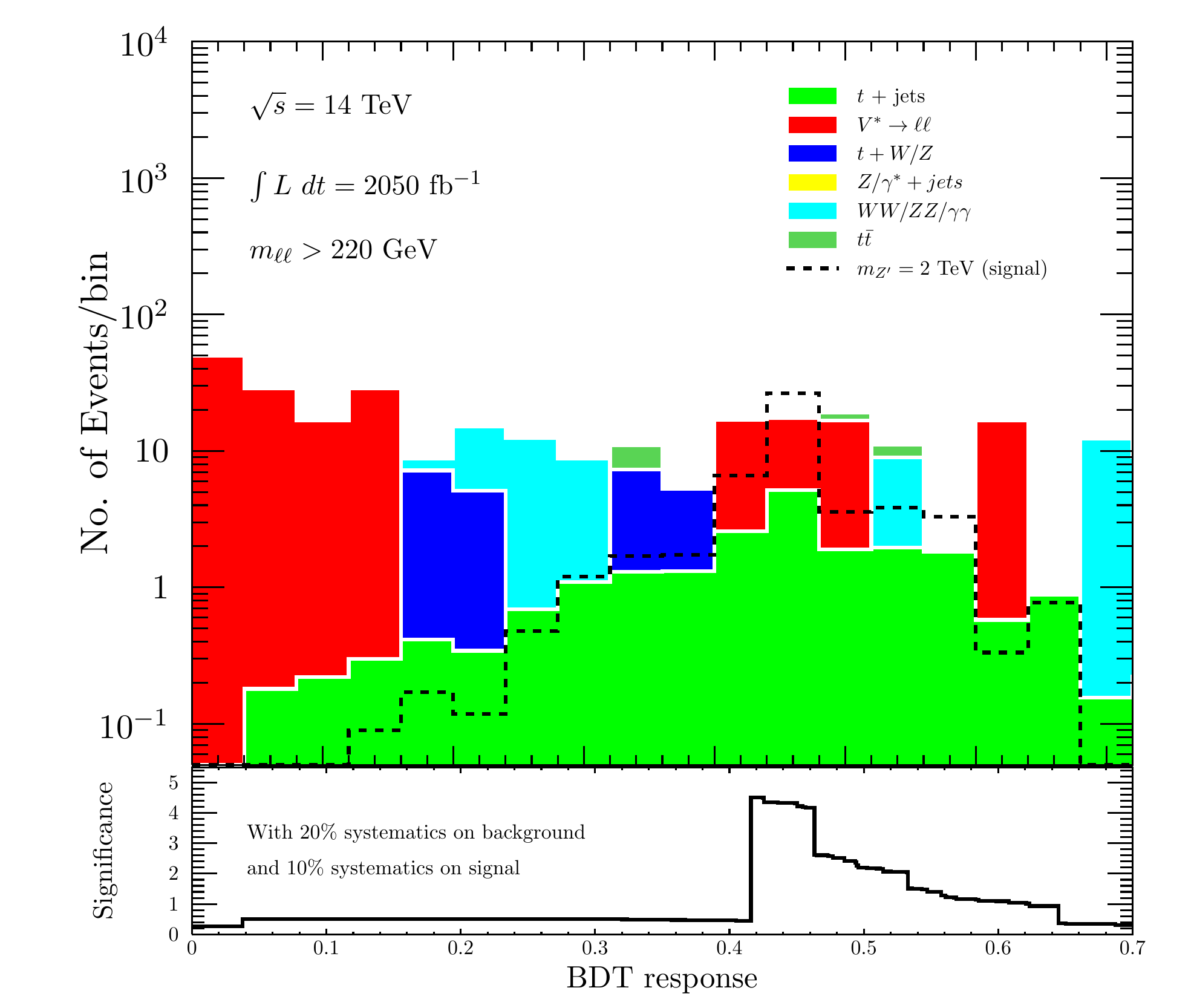}
\caption{Signal and background event distribution in the `BDT response' variable obtained after training and testing the BDT on the background and signal samples. The model points are the same as in Fig.~\ref{cutcount}.
The bottom plots show the significance given by Eq.~(\ref{merit}) as a result of applying cuts on the `BDT response'. Here, both model points are discoverable at the HL-LHC.}
\label{fig:bdta}
\end{figure}

We show in Fig.~\ref{fig:bdta} the distribution of signal and background events in the new BDT variable for the two benchmarks of choice. The lower panels indicate the effect of cuts on the significance defined in Eq.~(\ref{merit}). One can see from the left panel that a cut on the BDT response $>0.3$ produces a $5\sigma$ significance for an integrated luminosity of 2200 fb$^{-1}$, while from the right panel a cut $>0.4$ is required for discovery at an integrated luminosity of 2050 fb$^{-1}$. Note here that additional cuts are required to arrive at the desired results. Along with the cut on the BDT response, we require $m_{\ell\ell}>500$ GeV and $E_{T1}>350$ GeV (left panel) and $m_{\ell\ell}>1400$ GeV and $p_{T2}>350$ GeV (right panel). 

\section{Detection of a Stueckelberg dark photon at the Forward Physics Facility \label{sec:forward} }

As the kinetic mixing coefficient takes on smaller values, a dark photon becomes a long-lived particle (LLP) which can still decay to the SM inside a detector after having traveled a certain distance away from its production vertex. In the sub-GeV regime and aside from direct production (for e.g. Bremsstrahlung~\cite{Feng:2017uoz} and Drell-Yan production~\cite{Berlin:2018jbm}), a dark photon can be produced from the decay of a SM particle such as a meson. Light mesons are copiously produced in the forward region at the LHC and therefore constitute an important tool to study BSM physics with dark photons as leading candidates. The Forward Physics Facility (FPF)~\cite{Anchordoqui:2021ghd,Feng:2022inv} is intended to host several experiments which are able to capture the multitude of particles near the beam line which are otherwise missed by the current LHC experiments. So the FPF will be suited for studying and possibly detecting such particles.  In this Section we discuss the sensitivity reach of forward detectors at HL-LHC and future colliders in terms of discovering a Stueckelberg dark photon. 

\begin{figure}[H]
\centering
\includegraphics[width=0.65\textwidth]{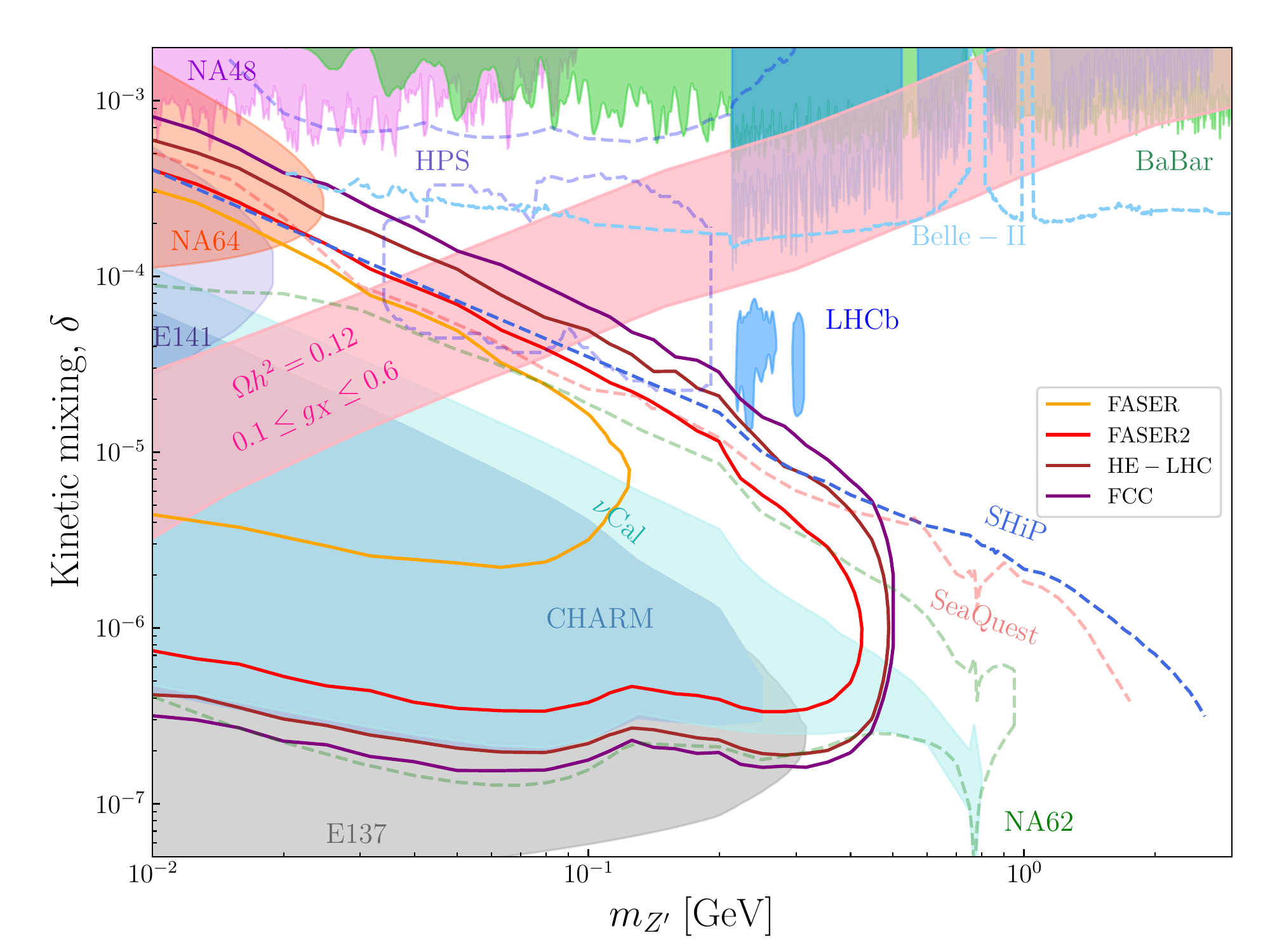}
\caption{The predicted sensitivity reach as determined by \code{FORESEE} for future forward detectors: FASER (orange line), FASER2 (red line), HE-LHC (brown line) and FCC (purple). Other experimental constraints are shown including CHARM~\cite{Tsai:2019buq}, $\nu$Cal~\cite{Tsai:2019buq,Blumlein:2011mv,Blumlein:2013cua}, E137~\cite{Andreas:2012mt}, E141~\cite{Riordan:1987aw}, NA64~\cite{NA64:2019auh}, NA48~\cite{NA482:2015wmo}, BaBar~\cite{BaBar:2014zli,BaBar:2017tiz}, HPS~\cite{HPS:2016jta}, LHCb~\cite{LHCb:2019vmc}, Belle-2~\cite{Belle-II:2018jsg}, SHiP~\cite{SHiP:2020vbd}, SeaQuest~\cite{Berlin:2018pwi,Tsai:2019buq} and NA62~\cite{Tsai:2019buq}. The pink band represents the relic density constraint consistent with Eq.~(\ref{Planck}). To ensure the dark photon will not decay to dark fermions, we set $m_D=0.6 m_{Z'}$.}
\label{fig:FORESEE}
\end{figure}

For our analysis, we consider the two production modes of a dark photon: direct production and production via meson decays. The latter requires a good understanding of the meson spectra, which has been studied and greatly improved over the years~\cite{LHCForwardPhysicsWorkingGroup:2016ote}. A new numerical package called \code{FORESEE}~\cite{Kling:2021fwx} (FORward Experiment SEnsitivity Estimator) allows users to implement their model and derive predictions on the sensitivity reach at future forward detectors. The package also provides the meson spectra which is necessary to determine the LLP flux generated from the decay of mesons. We implement our Stueckelberg dark photon model in \code{FORESEE} and provide the long-lived dark photon lifetime, its production rates and its decay branching ratios with the latter estimated using \code{DarkCast} (see Fig.~\ref{fig:br}). After taking into account the detector geometry and acceptance cuts, the number of surviving signal events are counted and used in \code{FORESEE} to draw the contours reflecting the sensitivity reach at forward detectors. We consider in this analysis the mass reach at FASER~\cite{FASER:2018ceo,FASER:2018eoc,FASER:2018bac}, which is already installed at the LHC beam line, FASER 2 which is planned for HL-LHC~\cite{CidVidal:2018eel}, as well as possible future detectors at HE-LHC~\cite{Todesco:2011np} and at the Future Circular Collider (FCC)~\cite{Mangano:2017tke}. The corresponding limits are shown in Fig.~\ref{fig:FORESEE}.

The projected limits are mainly derived from the dark photon decay channel $Z'\to e^+e^-$ whose branching ratio becomes progressively smaller for larger $Z'$ mass as hadronic decays become more favorable. This means one can no longer reach a dark photon mass larger than $\sim 0.5$ GeV when investigating dilepton final states. As one can see that there is a major gain in the reach along the kinetic mixing axis when going from FASER to the other future detectors. However, the gain is modest if one compares FASER2, HE-LHC and FCC. The same is true along the mass axis, where the future detectors can probe larger masses and major improvement from FASER is observed. The pink region is the part of the parameter space where the DM relic density is satisfied, i.e., consistent with Eq.~(\ref{Planck}). Several phenomenological work has been put forth regarding light $Z'$ at FASER~\cite{Asai:2022zxw,Cheung:2021tmx,Cheung:2022kjd} and heavy $Z'$ at future colliders~\cite{Das:2021esm}.

\section{Conclusion \label{sec:conclusion}}

In this work we have investigated the current constraints from collider experiments as well as from DM direct and indirect detection experiments on a well motivated extension of the electroweak sector of the Standard Model with a $U(1)_X$ gauge group. The $U(1)_X$ belongs to the hidden sector which also contains matter, that we assume to be a Dirac fermion field.  Although the hidden sector is neutral relative to the Standard Model gauge group, it can still communicate with the visible sector via kinetic mixing and via \stu mass mixing. We discussed the  kinetic energy in the mass diagonal basis, where one finds a massive dark photon or dark $Z'$
which can interact with the quarks and leptons in the visible sector with mixings characterized by the kinetic and mass mixing parameters. However, the dark photon has regular size couplings with the dark fermion allowing for a dark freeze-out to happen in the hidden sector, which generates the desired relic density for the dark fermions. In the analysis we have recast all the relevant constraints for a wide range of $Z'$ masses: above and below the $Z$ boson pole mass. For each case we considered both visible and dark decays of  $Z'$, where our analysis shows that the model parameter space is severely constrained for the case when a $Z'$ decays to DM. The reason is that the region with a large kinetic mixing is already excluded by the LHC and LEP and this region is important to deplete the DM abundance via $D\bar D\to f\bar f$ when the latter process is the only active one near freeze-out. However, for heavier DM masses, i.e., for $m_{Z'}>2m_D$, the dark photon can no longer decay to DM, and the process $D\bar D\to Z'Z'$ becomes the dominant channel for DM annihilation. This process is proportional to $g_X^4$ and so a large enough $g_X$ can weaken the relic density constraint thus opening up the available parameter space. Note that an increase in $g_X$ leads to more severe constraints from Xenon1T, LZ and Fermi-LAT. However, this is compensated by a small $f_{\rm DM}$ and a competition between $f_{\rm DM}$ and $g_X$ can go either way. In our analysis we see a slight increase in these constraints, but they are well tamed in such a way that parts of the model parameter space remain viable. We have also shown that unexplored regions of the parameter space can be accessible at HL-LHC and in forward detectors at the LHC and at future colliders. For the HL-LHC, we have carried out a detailed analysis for a potential discovery of a TeV mass scale $Z'$ and showed that one can observe a $5\sigma$ excess for a 1 TeV or 2 TeV $Z'$ using BDTs. Furthermore, we demonstrated the sensitivity reach of forward detectors for a sub-GeV dark photon at the LHC (FASER) and at future colliders (HE-LHC and FCC).  

Our analysis shows that the presence of a thermal hidden sector can weaken the current constraints on 
dark  $Z'$ models while the available regions of the model parameter space remain within reach of standard LHC searches as well as at forward detectors. \\

 \noindent
{\bf Acknowledgments:}  
The research of AA and MK was supported by the BMBF under contract 05P21PMCAA and by the DFG through the Research Training Network 2149 ``Strong and Weak Interactions - from
Hadrons to Dark matter" and grant KL 1266/10-1, while the research of PN and ZYW was supported in part by the NSF Grant PHY-2209903. The research of MMA was supported by the French Agence Nationale de la Recherche (ANR) grant no.\ ANR-21-CE31-0023 (PRCI SLDNP) and by the National Science Centre, Poland, under research grant 2017/26/E/ST2/00135. MK\ thanks the School of Physics at the University of New South Wales in Sydney, Australia for its hospitality and financial support through the Gordon Godfrey visitors program.

\appendix

\section{Rotation angles and dark matter couplings\label{app:A}}

Further details of the analysis presented in the main body of the paper
are given in this Appendix. As noted in section \ref{sec:model}, we have mixing of three vector bosons: $C^\mu, B^\mu, A^\mu_3$
which leads to diagonalization of both the kinetic energy matrix 
and the  mass matrix involving the fields.  In general this results in
a diagonalization of a $3\times 3$ vector boson mass-square matrix 
${\cal M}^2$ 
which, however, is symmetric and can be diagonalized by an orthogonal 
transformation ${\cal R}$ so that 
\begin{align}
\mathcal{R}^T\mathcal{M}^2 \mathcal {R}= \text{diag}(m^2_{Z'},  m_Z^2, 0),
\end{align}
 where $\mathcal{M}^2$ is defined by Eq.~(21) of~\cite{Feldman:2007wj}.
 The three Euler angles $\theta,\phi,\psi$ are given by 
\begin{equation}
 \tan\phi=-\sinh{\delta}, ~~~ \tan\theta=\frac{g_Y}{g_2}\cosh{\delta}\cos\phi,
~\tan2\psi=\frac{2 m^2_Z\sin\theta\tan\phi}{m^2_{Z'}-m^2_Z+(m^2_{Z'}+m^2_Z-m^2_W)\tan^2\phi},
 \label{hid-exact}
\end{equation} 
with the assumption of zero mass mixing.
Thus, the  couplings $g_{Z}, g_{\gamma}$ and $g_{Z'}$ that appear  in   $\Delta\mathcal{L}_{\rm int}$ (see Eq.~(\ref{D-darkphoton}))
are given by 
\begin{equation}
g_Z = g_X Q_X (\mathcal{R}_{12}- s_{\delta} \mathcal{R}_{22}), 
  ~g_{\gamma}= g_X Q_X (\mathcal{R}_{13}- s_{\delta} \mathcal{R}_{23}), 
 ~ g_{Z'}= g_X Q_X (\mathcal{R}_{11}- s_{\delta} \mathcal{R}_{21}).
\end{equation}

\section{More on the exclusion plots from \code{Contur}}\label{app:C}

In this appendix, we give the exclusion plots from precision measurements of the SM obtained using \code{Contur}. The plots are drawn for the parameters $\delta$ and $M_1$ over which the scan is made. The limits in these plots are then converted to the kinetic mixing-$m_{Z'}$ plane using $m_{Z'}^2=(q\pm p)/2$ and Eqs.~(\ref{p-eq}) and~(\ref{q-eq}). Let us begin by explaining the different data pools used by \code{Contur}.

An event that passes the cuts of a specific measurement can also be accepted in measurements that share similar final states. In order to avoid multiple counting of such events, and due to the lack of information about the correlations between different measurements, the analyses in \code{Rivet} are grouped into orthogonal pools based on three criteria: the experiment that conducted the measurement, the center of mass energy, and the considered final state. For each pool of analyses, a likelihood is built for every distribution taking the correlation between its bins\footnote{Note that this can only be done if the correlation information is provided by the experiment. If not, \code{Contur} will only consider the most sensitive bin of the histogram.} into account. The likelihoods of the orthogonal histograms within the pool are then combined. Finally, \code{Contur} constructs the total likelihood by combining the likelihoods of the different pools. 

\begin{figure}[H]
\centering
\includegraphics[width=0.495\textwidth]{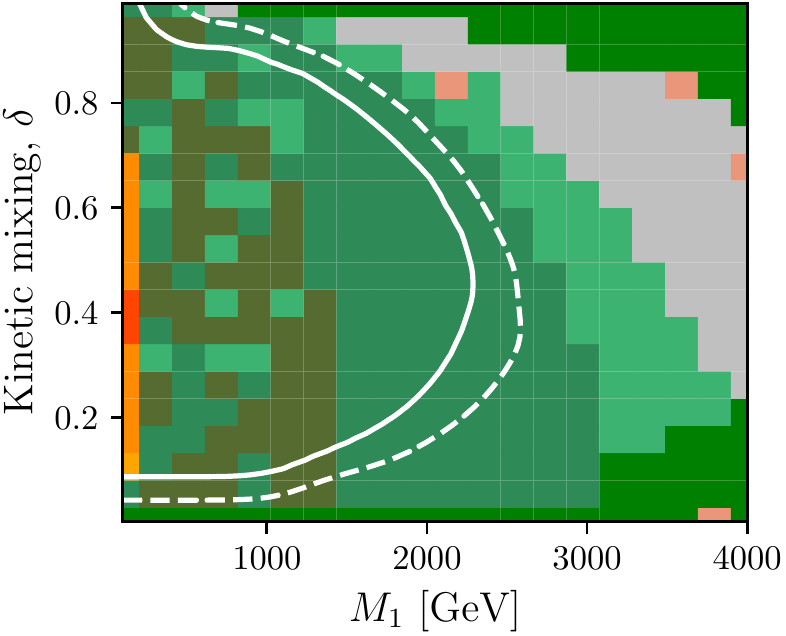}
\caption{The breakdown of \code{Contur}'s most sensitive analysis pool for each scan point in the heavy $Z'$ case. The solid (dashed) white line corresponds to the 95\% (68\%) CL exclusion on the kinetic mixing $\delta$ versus the $M_{1}$ parameter. Here $m_D =2m_{Z'}$ and so the limits correspond to the case of $Z'$ decaying to SM fermions only. }
\label{heavy-M1-no-decay}
{\small \begin{tabular}{lll}
    \swatch{green}~ATLAS \met{}+jet &
    \swatch{seagreen}~CMS high-mass Drell-Yan $\ell\ell$ &
    \swatch{orangered}~ATLAS $ee$+jet \\
    \swatch{orange}~ATLAS $\ell\ell$+jet &
    \swatch{darkolivegreen}~ATLAS high-mass Drell-Yan $\ell\ell$ &
    \swatch{mediumseagreen}~ATLAS $\ell\ell\gamma$ \\
    \swatch{darksalmon}~CMS $\mu\mu$+jet &
    \swatch{darkorange}~ATLAS $\mu\mu$+jet &
    \swatch{silver}~ATLAS jets 
\end{tabular} }
\end{figure}

\begin{figure}[H]
\centering
\includegraphics[width=0.495\textwidth]{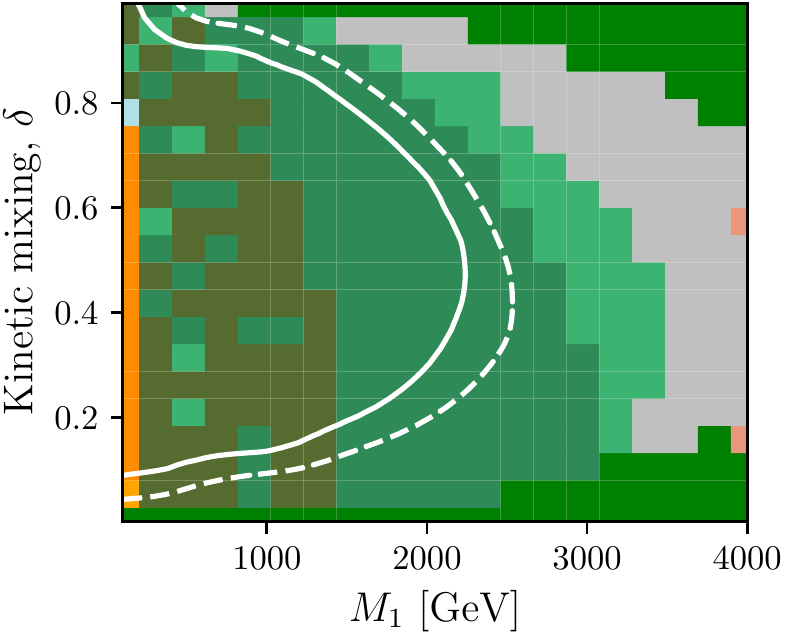}
\includegraphics[width=0.495\textwidth]{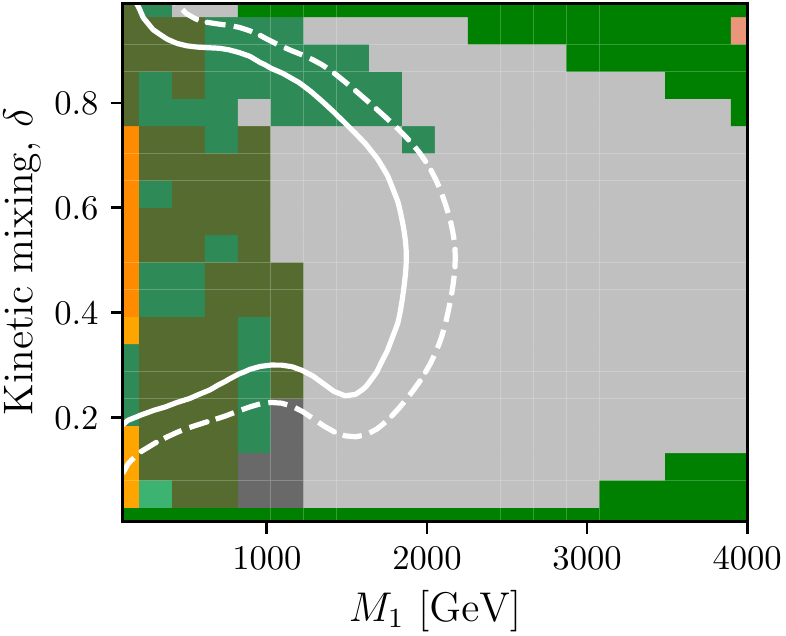}
\caption{The 95\% CL (solid) and the 68\% CL (dashed line) exclusions on the kinetic mixing $\delta$ versus the $M_{1}$ parameter for the case of heavy $Z'$ invisible decay to DM fermions and for different values of $g_X$. The left panel corresponds to $g_X=0.1$ while the right panel corresponds to $g_X=1.0$. We take $m_D=m_{Z'}/4$. The colored bins correspond to the pool of analyses giving the dominant exclusion.}
\label{fig3-heavy-M1a}
{\small \begin{tabular}{lll}
        \swatch{green}~ATLAS \met{}+jet &
        \swatch{darkolivegreen}~ATLAS high-mass Drell-Yan $\ell\ell$ &
        \swatch{orange}~ATLAS $\ell\ell$+jet \\
        \swatch{darkorange}~ATLAS $\mu\mu$+jet &
        \swatch{silver}~ATLAS jets &
        \swatch{mediumseagreen}~ATLAS $\ell\ell\gamma$ \\
        \swatch{darksalmon}~CMS $\mu\mu$+jet &
        \swatch{dimgrey}~CMS jets &
        \swatch{seagreen}~CMS high-mass Drell-Yan $\ell\ell$\\
        \swatch{powderblue}~CMS $\ell$+\met{}+jet 
\end{tabular} }
\end{figure}

\begin{figure}[H]
\centering
\includegraphics[width=0.45\textwidth]{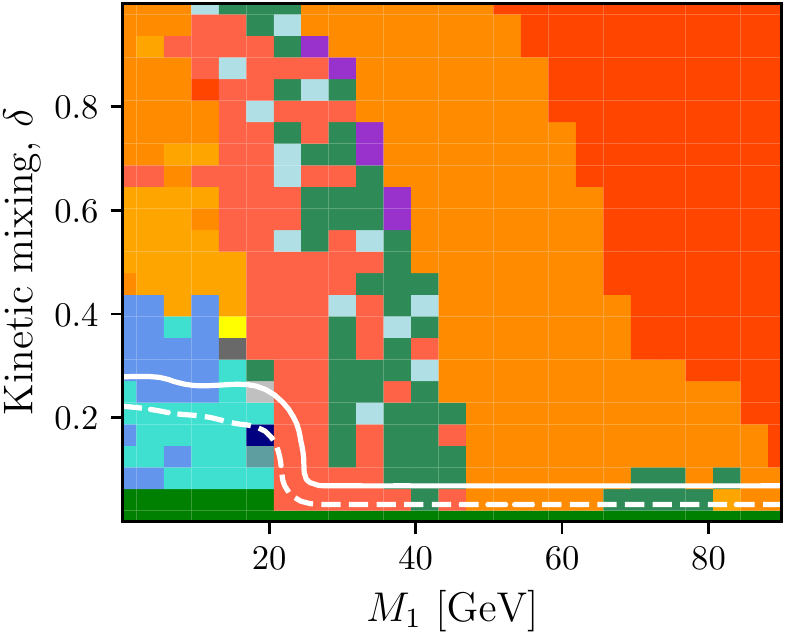}
\caption{The 95\% CL (solid) and the 68\% CL (dashed line) exclusions on the kinetic mixing $\delta$ versus the $M_1$ parameter . Here $m_D=2m_{Z'}$ and so the limits correspond to the case of light $Z'$ decaying to SM fermions only. The scan on $\delta$ is done in the range $10^{-5}$ to $1$ and the colored bins correspond to the analyses pool giving the dominant contribution.}
\label{fig1-light-M1a}
{\small \begin{tabular}{lll}
        \swatch{yellow}~ATLAS $\gamma$ &
        \swatch{tomato}~ATLAS low-mass Drell-Yan $\ell\ell$ &
        \swatch{orangered}~ATLAS $ee$+jet \\
        \swatch{cornflowerblue}~ATLAS $\ell_1\ell_2$+\met{} &
        \swatch{seagreen}~CMS high-mass Drell-Yan $\ell\ell$ &
        \swatch{cadetblue}~ATLAS $e$+\met{}+jet \\
        \swatch{green}~ATLAS \met{}+jet &
        \swatch{powderblue}~CMS $\ell$+\met{}+jet &
        \swatch{dimgrey}~CMS jets \\
        \swatch{silver}~ATLAS jets &
        \swatch{orange}~ATLAS $\ell\ell$+jet &
        \swatch{darkorange}~ATLAS $\mu\mu$+jet \\
        \swatch{turquoise}~ATLAS $\ell_1\ell_2$+\met{}+jet &
        \swatch{darkorchid}~LHCb $\ell$+jet &
        \swatch{navy}~ATLAS $\mu$+\met{}+jet
\end{tabular} }
\end{figure}

\begin{figure}[H]
\centering
\includegraphics[width=0.45\textwidth]{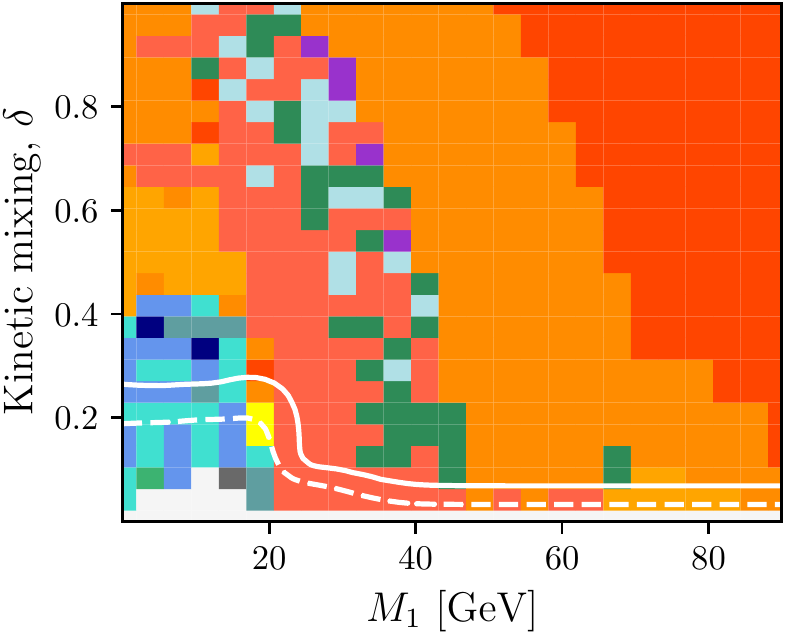}
\includegraphics[width=0.45\textwidth]{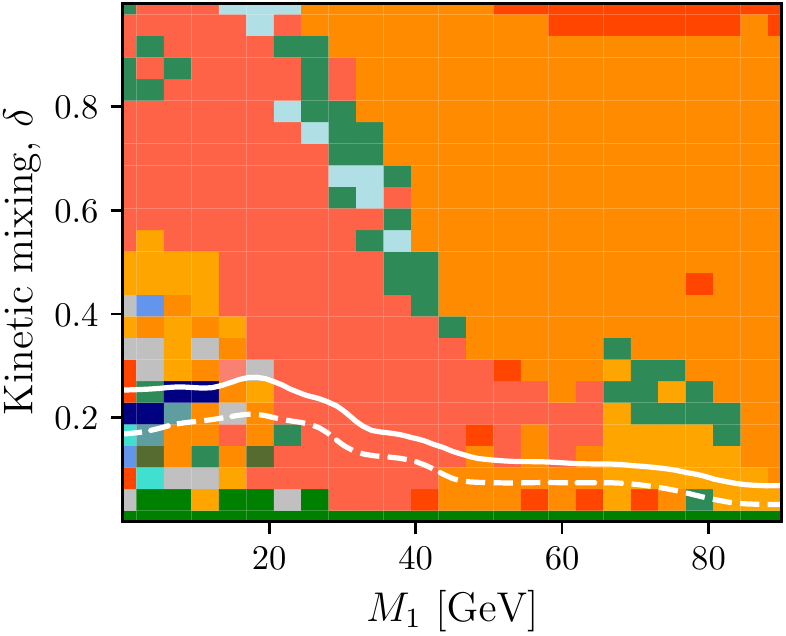}
\caption{The 95\% CL (solid) and the 68\% CL (dashed line) exclusions on the kinetic mixing $\delta$ versus the $M_{1}$ parameter for the case of light $Z'$ invisible decay to DM fermions with $m_D=m_{Z'}/4$ and for $g_X=0.1$ (left panel) and $g_X=1.0$ (right panel). The color-shading scheme specifies which SM measurement has the dominant exclusion.}
\label{fig3-light-M1a}
{\small \begin{tabular}{lll}
        \swatch{darkorange}~ATLAS $\mu\mu$+jet &
        \swatch{powderblue}~CMS $\ell$+\met{}+jet &
        \swatch{yellow}~ATLAS $\gamma$ \\
        \swatch{orangered}~ATLAS $ee$+jet &
        \swatch{green}~ATLAS \met{}+jet &
        \swatch{silver}~ATLAS jets \\
        \swatch{cornflowerblue}~ATLAS $\ell_1\ell_2$+\met{} &
        \swatch{darkolivegreen}~ATLAS high-mass Drell-Yan $\ell\ell$ &
        \swatch{salmon}~CMS $\ell\ell$+jet \\
        \swatch{turquoise}~ATLAS $\ell_1\ell_2$+\met{}+jet &
        \swatch{navy}~ATLAS $\mu$+\met{}+jet &
        \swatch{orange}~ATLAS $\ell\ell$+jet \\
        \swatch{tomato}~ATLAS low-mass Drell-Yan $\ell\ell$& 
        \swatch{darkorchid}~LHCb $\ell$+jet &
        \swatch{dimgrey}~CMS jets \\
        \swatch{seagreen}~CMS high-mass Drell-Yan $\ell\ell$ &
        \swatch{cadetblue}~ATLAS $e$+\met{}+jet &
        \swatch{mediumseagreen}~ATLAS $\ell\ell\gamma$
\end{tabular} }
\end{figure}

The plots in this Appendix show the most sensitive pool at each point of the parameter space for the different scenarios that we consider in this paper, i.e.\ heavy $Z'$ bosons decaying to SM fermions only (Fig.\ \ref{heavy-M1-no-decay}) or also DM (Fig.\ \ref{fig3-heavy-M1a}) as well as light $Z'$ bosons decaying to SM fermions only (Fig.\ \ref{fig1-light-M1a}) or also DM (Fig.\ \ref{fig3-light-M1a}).

\vspace{-0.3cm}

\end{document}